\documentclass[english,aps,twocolumn,floatfix, showpacs, amsfonts, amssymb,prb]{revtex4}
\usepackage{epsfig, graphicx,psfrag,amsmath,amssymb,float}
\usepackage[T1]{fontenc}
\usepackage[latin9]{inputenc}
\usepackage{amsmath}
\usepackage{amssymb}
\usepackage{amscd}
\usepackage{bm}
\usepackage{psfrag}
\usepackage{babel}
\usepackage{graphicx}
\usepackage[dvips]{color}
\usepackage{dsfont}

\begin{document}

\title{Universal low-temperature crossover in two-channel Kondo models}

\author{Andrew K. Mitchell and Eran Sela}

\affiliation{Institute for Theoretical Physics, University of Cologne, 50937 Cologne, Germany}


\begin{abstract}
An exact expression is derived for the electron Green function in
two-channel Kondo models with one and two impurities, describing the
crossover from non-Fermi liquid (NFL) behavior at intermediate
temperatures to standard Fermi liquid (FL) physics at low temperatures.
Symmetry-breaking perturbations generically present in experiment
ensure the standard low-energy FL description, but
the full crossover is wholly characteristic of the
unstable NFL state.
Distinctive conductance lineshapes in quantum dot devices should result.
We exploit a connection between this crossover and one occurring in a
classical boundary Ising model to calculate
real-space electron densities at finite temperature.
The single universal finite-temperature Green function is then extracted
by inverting the integral transformation relating these Friedel
oscillations to the t matrix. Excellent agreement is demonstrated
between exact results and full numerical renormalization group 
calculations.
\end{abstract}

\date{\today}

\maketitle


 \section{Introduction and Physical picture}
The full power of the renormalization group (RG) concept is perhaps
most clearly seen in its application to quantum impurity
systems.\cite{hewson} The classic paradigm is the Kondo
model,\cite{kondo} being the simplest to capture the fundamental
physics associated with all quantum impurity models:  
universal RG flow from an unstable fixed 
point (FP) to a stable one on successive reduction of the temperature
or energy scale. The Kondo model describes a single local
spin-$\tfrac{1}{2}$ `impurity', coupled by antiferromagnetic exchange
to a single channel of noninteracting conduction electrons. Here,
perturbative scaling arguments\cite{anderson} indicate an RG flow from
a high-energy unstable `free fermion' FP (describing a free impurity
decoupled from a free conduction band), to a low-energy stable `strong
coupling' FP (where the impurity is
screened by conduction electrons via formation of a `Kondo
singlet'). This RG flow is characterized by a scaling invariant ---
the Kondo temperature $T_K$ --- which sets the crossover energy
scale. But analysis of the
crossover itself goes beyond simple scaling ideas and the conventional
RG picture. Wilson's numerical renormalization group\cite{wilson}
(NRG) allows an exact nonperturbative calculation of certain thermodynamic
and dynamical quantities which show the crossover (for a review,
see Ref.~\onlinecite{nrg:rev}).
Universal scaling of all physical quantities in terms of the crossover
scale $T_K$ confirms the basic RG structure of the problem.

However, a different RG flow occurs when the impurity is coupled to
two or more independent conduction channels.\cite{NozieresBlandin}
In this multichannel Kondo model, the frustration inherent when
several channels compete to screen the impurity spin
renders the strong coupling FP unstable. A third FP at
\emph{intermediate} coupling\cite{NozieresBlandin} then dictates the
low-energy physics. This FP exhibits non-Fermi liquid (NFL) behavior,
including notably a residual entropy\cite{s} of
$\tfrac{1}{2}k_B\ln(2)$ in the two-channel Kondo (2CK) model. The
crossover from the free fermion FP to the 2CK FP has been the focus of
much theoretical attention. In particular,
solution of the model using the Bethe ansatz yields the exact
crossover behavior of thermodynamic quantities;\cite{s} while NRG has
been used to calculate thermodynamics\cite{2cknrgtd} and
dynamics\cite{Toth,Toth2} numerically.
It was also shown recently that this 2CK physics
can arise in odd-membered quantum dot rings\cite{akm:rings} and
chains,\cite{akm:oddimp} and in quantum box systems.\cite{matveev_box,lehur_box,lsa_box,kek_box}

Indeed, the same type of NFL
behavior\cite{gan,Zarand2006,akm_2ckin2ik} arises in the
two-impurity Kondo (2IK) model.\cite{jones} The tendency to form a
trivial local singlet state is favored by an exchange coupling acting
directly between the impurities; while the coupling of each impurity
to its own conduction channel favors separate single-channel Kondo
screening. The resulting competition gives rise to a critical
point\cite{jones} that is identical to that of the 2CK model with
additional potential scattering.\cite{akm_2ckin2ik}
2IK physics is also expected to appear in certain double quantum dot
systems,\cite{2iam_box} and other even-impurity
chains.\cite{Zarand2006}

A description of the NFL FPs of such two-channel models in terms of an
effective boundary conformal field theory (CFT)
shows that the operator controlling the FP has an anomalous scaling
dimension.\cite{CFT2CK,CFT2IKM} This implies unconventional energy/temperature dependences of
physical quantities such as conductance $G_c(V,T)$, measured as a
function of bias voltage $V$ and temperature $T$. In the 2CK device
of Ref.~\onlinecite{Potok07}, $\sqrt{V/T_K}$  and $\sqrt{T/T_K}$
corrections to the NFL FP conductance predicted by CFT were directly
observed in experiment. Similar signatures are
expected\cite{Zarand2006,2iam_box} in the channel-asymmetric 2IK model;
although leading linear behavior emerges in the symmetric
2IK.\cite{akm_2ckin2ik} This behavior is of course in marked
contrast to $(V/T_K)^2$ and $(T/T_K)^2$ Fermi liquid (FL) behavior
obtained ubiquitously in the single-channel case.\cite{1ckexpt}

\begin{figure}[t]
\begin{center}
\includegraphics*[width=87mm]{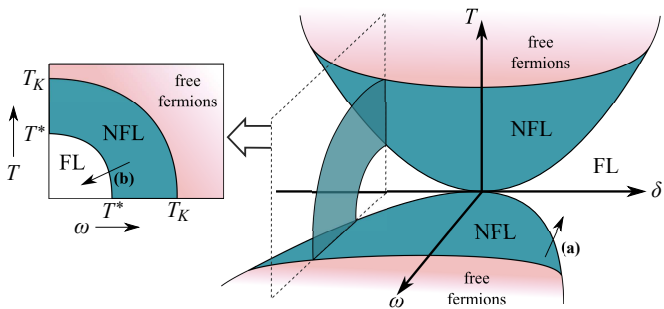}
\caption{\label{fig:pd}
  Schematic phase diagram for the 2CK and 2IK
  models, as a function of temperature $T$, external energy scale
  $\omega$, and symmetry-breaking perturbation strength $\delta$. The three FPs
  of each model give rise to three distinct regimes: free fermion, NFL
  and FL. We considered the NFL to FL crossover at $T=0$ in
  Ref.~\onlinecite{SelaMitchellFritz}, indicated by arrow
  (a). Here we generalize the results to finite temperature, arrow (b).
}
\end{center}
\end{figure}

The NFL FP itself (and the crossover to it) has now been rather
well studied. However, NFL physics is extremely delicate:
various symmetry-breaking perturbations destabilize the NFL
FP\cite{CFT2CK,CFT2IKM} and generate a new crossover scale $T^*$.
At $T=0$, the impurities are thus completely screened and all
residual entropy is quenched.
Indeed, regular FL behavior,\cite{hewson} including the standard
$(V/T^*)^2$ and $(T/T^*)^2$ corrections to conductance, must appear
at low temperatures $T\ll T^{*}$ and energies $V\ll T^*$. Therefore,
no evidence of nascent NFL physics can manifest in the immediate
vicinity of the FL FP.
Only on fine-tuning the perturbation strength $\delta\rightarrow 0$ to
the critical point so that $T^*\rightarrow 0$ does one obtain NFL
physics on the lowest energy scales.

But RG analysis in the vicinity of the free fermion, NFL and FL FPs
implies two successive crossovers, with $T_K$ setting the energy scale
for flow to the NFL FP, and $T^*$ characterizing flow away from
it. Even in the FL phase away from the critical point (which is
the generic case relevant to experiment), NFL behavior can be observed
at \emph{higher} temperatures and energies, provided there is good
scale separation $T^*\ll T_K$ (see Fig.~\ref{fig:pd} for a schematic
phase diagram). In this case, conductance $G_c(V,T)$ through the 2CK
quantum dot device of Ref.~\onlinecite{Potok07}, or proposed 2IK
devices,\cite{Zarand2006,akm_2ckin2ik} should exhibit a clean NFL to
FL crossover.

In Ref.~\onlinecite{SelaMitchellFritz}, we considered this conductance
crossover at $T=0$ as a function of bias $V$, corresponding to
the crossover labelled by arrow (a) in Fig.~\ref{fig:pd} (and $V$
here playing the role of the external energy scale $\omega$).
It was shown\cite{SelaMitchellFritz} that the full crossover is
wholly characteristic of the high-energy NFL
state. The $T=0$ crossover is expected to describe the behavior at
very low temperatures. From a scaling perspective, RG flow is cut
off on the energy scale $V\sim T$, so $G_c(V,T)\simeq
G_c(V,0)$ for $T\ll T^{*}$ since there are no further crossovers
below $T^{*}$.

By contrast, at higher temperatures $T\gg T^*$ ($\ll T_K$), no
evidence of the NFL to FL crossover will be observed (see
Fig.~\ref{fig:pd}), and only the NFL FP is probed.
Indeed, this is the likely scenario in the experiment of
Ref.~\onlinecite{Potok07}: rather than tuning to the critical point
$\delta=0$, signatures of the true FL ground state are simply washed
out by temperature.
But the behavior as a full function of temperature $T$ and energy
scale $\omega$ for finite perturbation strength $\delta$ is much more subtle,
and naturally strengthens connection to experiment.
Exploring the third temperature axis in Fig.~\ref{fig:pd}, and
considering the resultant NFL to FL crossover [eg. arrow (b)], is thus
the focus of the present work.


In this paper we combine Abelian bosonization
methods\cite{emery,gogolin,gan} with the powerful machinery of
CFT\cite{CFT2CK,CFT2IKM} to obtain an exact description of the NFL to
FL crossover in two-channel Kondo models. In particular, we calculate
the full electron Green function at finite temperature, from which
conductance follows.\cite{meir,asymmetric}
The field theoretic description links the 2IK model
with a classical Ising model on a semi-infinite plane.\cite{CFT2IKM}
Application of a boundary magnetic field $h$ in this boundary Ising model
(BIM) results in RG flow from an unstable FP with free boundary
condition $h=0$ to a stable FP with fixed boundary condition
$h\rightarrow \pm \infty$.\cite{cardy,Cardy91} This RG flow due to $h$ is
identical to that occurring between NFL and FL FPs in the 2IK model due
to a small perturbation $\delta$.\cite{CFT2IKM}  Indeed, an emergent
symmetry of the NFL FP in the 2IK model,\cite{CFT2IKM}  together with
the common CFT description of 2IK and 2CK
models,\cite{CFT2CK,CFT2IKM,akm_2ckin2ik}
implies the existence of a single universal NFL to FL crossover
function for both models, resulting from any combination of relevant
perturbations.\cite{SelaMitchellFritz}

Exact results\cite{cz,Leclair} for the BIM are the source of our
solution, which becomes exact when there is good scale separation
$T^{*} \ll T_K$, as sought experimentally.

The exact crossover Green function at $T=0$ was calculated in
Ref.~\onlinecite{SelaMitchellFritz} by exploiting the above
connection. However, ambiguities appear at finite temperature 
which prevent straightforward generalization of those results. Thus we
take a different route here: the BIM solution is used to
calculate real-space Friedel oscillations around the impurities at finite
temperature, which are themselves related by integral
transformation\cite{akm:realspace} to the Green function. The
problematic analytic continuation is avoided in this way.


The paper is organized as follows. In Sec.~\ref{model} we
introduce the 2CK and 2IK models, together with representative
symmetry-breaking perturbations that generate the NFL to FL
crossover. We then present and discuss our main results for the exact
finite-temperature Green function along the crossover. The
corresponding conductance crossover for quantum dot
systems that might realize 2CK or 2IK physics is then calculated. In
Secs.~\ref{2ikgf}--\ref{gengf} we derive the analytic results. First
we consider the 2IK model at $T=0$ with a single detuning
perturbation. In Sec.~\ref{2ikgf} we calculate the resulting crossover
Green function, exploiting the analogy to the BIM. In
Sec.~\ref{finiteTderiv} we extend the calculation to finite
temperatures, extracting the desired t matrix from Friedel
oscillations. The results are generalized to the 2CK model
in Sec.~\ref{2ckgen} and to an arbitrary combination of perturbations
in Sec.~\ref{gengf}. Exact results are compared with 
finite-temperature NRG calculations in Sec.~\ref{nrgcomp}. Other
quantities showing the crossover such as entropy and 
nonequilibrium transport are then briefly considered in
Sec.~\ref{other}. The paper concludes with a general discussion in
Sec.~\ref{concs}, and details of certain calculations can be found in
the appendices.


\section{Models and Results}
\label{model}

We consider the standard 2CK and 2IK models,
\begin{align}
\label{2ck}
H_{2CK} =H_0+&J  \vec{S} \cdot (\vec{s}_{0L} + \vec{s}_{0R})  +\delta
H_{2CK},\\
\label{2ik}
H_{2IK} =H_0+&J(  \vec{S}_L \cdot \vec{s}_{0L} + \vec{S}_R
\cdot \vec{s}_{0R} ) +K \vec{S}_L \cdot \vec{S}_R + \delta H_{2IK}, \nonumber \\
\end{align}
where $H_{0}=\sum_{\alpha,k} \epsilon_{k}^{\phantom{\dagger}} \psi_{k\sigma\alpha}^{\dagger}
 \psi^{\phantom{\dagger}}_{k \sigma\alpha}$ describes two free
conduction electron channels $\alpha=L/R$, with spin density
$\vec{s}_{0\alpha}=\sum_{\sigma\sigma'}\psi_{0\sigma\alpha}^{\dagger}
(\tfrac{1}{2}\vec{\sigma}_{\sigma\sigma'})\psi^{\phantom{\dagger}}_{0
  \sigma' \alpha}$ (and $\psi_{0\sigma\alpha}^{\dagger}=
\sum_{k}\psi_{k\sigma\alpha}^{\dagger}$) coupled to one spin-$\tfrac{1}{2}$ impurity
$\vec{S}$ (2CK) or two impurity spins $\vec{S}_{L,R}$ (2IK).
For $\delta H_{2CK}=0$, the NFL ground state
of $H_{2CK}$ is described by the 2CK FP. Likewise, a critical
inter-impurity coupling $K_c$ can be found such that the ground state
of $H_{2IK}$ is again a NFL,\cite{jones} and is similarly described by
the 2CK FP for $\delta H_{2IK}=0$.\cite{gan,Zarand2006,akm_2ckin2ik}

Relevant perturbations to the above models (embodied by $\delta
H_{2CK}$ and $\delta H_{2IK}$) are those that destabilize
the NFL FP, and result in a FL ground state. A new scale $T^{*}$ is
thus generated, characterizing RG flow from NFL to FL FPs.
The relevance of such perturbations can be traced to the breaking of
certain symmetries,\cite{CFT2CK,CFT2IKM}
such as parity or particle-hole symmetries. In fact,
there are many possible perturbations to the 2CK and 2IK
models; but two perturbations may be considered `equivalent' if they
break the same underlying symmetry --- and their effect on the
low-energy physics will be identical.\cite{CFT2CK,CFT2IKM}

For concreteness, we consider now the simplest perturbations which
exemplify such symmetry-breaking, and which in combination generate
all possible NFL to FL crossovers at low
energies/temperatures. Specifically,
\begin{equation}
\label{delta2CK}
\delta H_{2CK} =\sum_{\ell = x,y,z} \Delta_\ell \sum_{\alpha,\beta}\sum_{\sigma\sigma'}
\psi_{0\sigma\alpha}^{\dagger} (\tfrac{1}{2}\vec{\sigma}_{\sigma\sigma'}
\tau_{\alpha\beta}^{\ell})\psi_{0 \sigma'\beta}  \cdot \vec{S} +
\vec{B} \cdot \vec{S},
\end{equation}
describes $L/R$ channel asymmetry in the 2CK model for $\Delta_z\ne
0$, while charge transfer between the leads is embodied in the
$\Delta_x$ and $\Delta_y$ components of the first term [here
$\vec{\tau} ~(\vec{\sigma})$ are the Pauli matrices in the channel (spin) sector]. The second
term describes a magnetic field acting on the impurity.
For the 2IK model, the critical point is destabilized by finite
$(K_c-K)$, and also through
\begin{equation}
\label{delta2IK}
\delta H_{2IK} = \sum_{\sigma}(V_{LR} \psi_{0\sigma L}^{\dagger} \psi^{\phantom{\dagger}}_{0 \sigma R} +
\text{H.c.}) + \vec{B}_s \cdot (\vec{S}_L - \vec{S}_R),
\end{equation}
where $V_{LR}$ describes electron tunneling between the leads
and $\vec{B}_s$ the application of a staggered magnetic field.
Channel anisotropy could also be included in the 2IK model, but the
critical point can always be recovered\cite{Zarand2006} on retuning
$K$. Similarly, spin-assisted
tunneling between channels $\sum_{\sigma} \psi_{0\sigma L}^{\dagger}
\psi^{\phantom{\dagger}}_{0 \sigma R} \vec{S}_L \cdot \vec{S}_R +
\text{H.c.}$ (as arises in a two-impurity Anderson
model\cite{DEL_2iam}) is expected to have the same destabilizing
effect as the $V_{LR}$ term in Eq.~\ref{delta2IK}, since they both
have the same symmetry at the NFL FP\cite{erratum} (although the
resulting crossover energy scales may themselves be
different\cite{DEL_2iam}). Thus we do not consider such 
perturbations explicitly here.


\subsection{Quantities of interest}
\label{condtmatrix}
\begin{figure}[t]
\begin{center}
\includegraphics*[width=75mm]{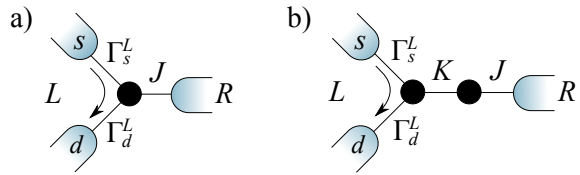}
\caption{\label{fig:condpic} Schematic illustration of possible 2CK
  (a) and 2IK (b) setups to measure conductance. The left lead is
  `split' into source and drain, allowing the resulting conductance
  through the attached impurity to be measured.}
\end{center}
\end{figure}

Signatures of the NFL to FL crossover on the scale of $T^{*}$ should
appear in all physical quantities. In 2CK or 2IK quantum dot devices
which could access this physics,\cite{Potok07,Zarand2006,akm_2ckin2ik}
the quantity of interest is the $dI/dV$ conductance
$G_c^{\alpha}(V,T)\equiv(2e^2 h^{-1}
G_0^{\alpha})\tilde{G}_c^{\alpha}(V,T)$ through channel $\alpha=L$ or
$R$. Here, $G^{\alpha}_0=4\Gamma^{\alpha}_s\Gamma^{\alpha}_d/
(\Gamma^{\alpha}_s+\Gamma^{\alpha}_d)^2$ describes the relative
strength of coupling to source and drain leads (see
Fig.~\ref{fig:condpic} for an illustration of the setup).
At zero-bias  $V=0$, the conductance is given exactly by,\cite{meir}
\begin{equation}
\label{zbcond}
\tilde{G}^{\alpha}_c(0,T)=\tfrac{1}{2}\sum_{\sigma}
\int^{\infty}_{-\infty}\textit{d}\omega\frac{-\partial
  f(\omega,T)}{\partial \omega} t_{\sigma\alpha}(\omega,T),
\end{equation}
where $f(\omega,T)=[e^{\omega/T}+1]^{-1}$ is the Fermi function, and
$t_{\sigma\alpha}(\omega,T)$ is the energy-resolved local density of states (or `spectrum'),
\begin{equation}
\label{specdef}
t_{\sigma\alpha}(\omega,T)=-\pi\nu~\text{Im}\mathcal{T}_{\sigma\alpha,\sigma\alpha}(\omega,T),
\end{equation}
itself related to the t matrix\cite{hewson}
$\mathcal{T}_{\sigma\alpha,\sigma\alpha}(\omega,T)$, describing
scattering of a $\sigma=\uparrow$ or $\downarrow$ electron within
channel $\alpha=L$ or $R$, with bare lead density of
states per spin $\nu$. Note that for equal hybridization to source
and drain leads, $\Gamma^{\alpha}_s=\Gamma^{\alpha}_d$, $G^{\alpha}_0=1$ is maximal;
while in the asymmetric limit $\Gamma^{\alpha}_s\ll\Gamma^{\alpha}_d$,
$G^{\alpha}_0\ll 1$. In the latter case, the weakly-coupled source
lead probes the system perturbatively, so the
system remains near equilibrium, even at finite bias $V>0$. The
resulting conductance is then simply,\cite{asymmetric}
\begin{equation}
\label{asymcond}
\tilde{G}^{\alpha}_c(V,T)=\tfrac{1}{2}\sum_{\sigma}
\int^{\infty}_{-\infty}\textit{d}\omega\frac{-\partial
  f(\omega-V,T)}{\partial \omega} t_{\sigma\alpha}(\omega,T),
\end{equation}
where $t_{\sigma \alpha}$ is the equilibrium (zero-bias) spectrum.

The t matrix itself must show signatures of the NFL to FL crossover
since scattering is purely inelastic at the NFL
FP,\cite{CFT2CK,CFT2IKM} but inelastic scattering must cease at
energies $\ll T^{*}$, where the impurity degrees of freedom are fully
quenched.\cite{borda} Thus the crossover also shows up in
conductance. Our goal here is to calculate the full crossover t matrix,
and hence conductance, at finite temperature
for the 2CK and 2IK models in the presence
of symmetry-breaking perturbations described by
Eqs.~\ref{2ck}--\ref{delta2IK}.


\subsection{Survey of results and discussion}
\label{results}

In the next sections we derive an exact expression for the desired t
matrix, describing the universal crossover from NFL to FL behavior
in the 2CK and the 2IK models at finite temperature --- and which as
such generalize the results of our previous work in
Ref.~\onlinecite{SelaMitchellFritz}. Here we pre-empt the full
derivation, and present our key results.

The NFL to FL crossover is characterized by a low-energy scale $T^{*}$
arising due to the presence of symmetry-breaking perturbations to the
2CK and 2IK models. It is given generically
by\cite{note:convention}
\begin{equation}
\label{tstar}
T^{*}=\lambda^2,
\end{equation}
where $\lambda^2= \sum_{j=1}^8 \lambda_j^2$. The eight contributions
correspond to relevant perturbations which have distinct symmetry at
the NFL FP. Two perturbations which have the same symmetry correspond
to the same $\lambda_j$.  The perturbations given in
Eqs.~\ref{delta2CK} and \ref{delta2IK} are classified viz,

\begin{table}[h]
\caption{Classification of perturbations}
\centering
\begin{tabular}{c c c}
\hline\hline
$\lambda_j$ &  ~~~~~2CK model~~~~~  & ~~~~~2IK model~~~~~ \\ [0.5ex]
\hline
$\lambda_1$       & $c_1\nu\Delta_z\sqrt{T_K}$        & $c_1(K_c-K)/\sqrt{T_K}$ \\
$\lambda_2$       & $c_1\nu\Delta_{x}\sqrt{T_K}$    & $c_V\text{Re}~\nu V_{LR}\sqrt{T_K}$ \\
$\lambda_3$       & $c_1\nu\Delta_{y}\sqrt{T_K}$    & $c_V\text{Im}~\nu V_{LR}\sqrt{T_K}$ \\
$\vec{\lambda}_B$ & $c_B \vec{B}/\sqrt{T_K}$      & $c_B \vec{B}_s/\sqrt{T_K}$ \\ [1ex]
\hline
\end{tabular}
\label{table:pert}
\end{table}
\noindent where $\vec{\lambda}_B\equiv \{\lambda_B^x,\lambda_B^y,\lambda_B^z \} =
 \{\lambda_4,\lambda_5,\lambda_6 \}$. The perturbations associated
 with coupling constants $\lambda_7$ and $\lambda_8$  do not conserve
 total charge,\cite{CFT2CK,CFT2IKM} and so are ignored here (although
 we note that such operators can be of importance, for example, in the
 context of strongly correlated superconductors\cite{supercond}).

$c_1, c_V, c_B = \mathcal{O}(1)$ are
fitting parameters\cite{note:convention}
which depend on the model and on $J$, and $T_K \propto e^{-\frac{1}{\nu J}}$
is the Kondo temperature, characterizing RG flow from the high-energy
free fermion FP to the NFL FP.\cite{akm_2ckin2ik} We do not discuss
the high-energy crossover in the present work.

The various perturbations described by Eqs.~\ref{delta2CK} and
\ref{delta2IK} describe very different physical processes --- but the
resulting crossover scale Eq.~\ref{tstar}, has a simple form due to an
\emph{emergent} $SO(8)$ symmetry of the effective NFL FP Hamiltonians,
as discussed in the following sections.

The main result of this paper is the NFL to FL crossover t matrix,
given by,
\begin{eqnarray}
\label{exactresult}
2 \pi i \nu \mathcal{T}_{\sigma \alpha,  \sigma' \alpha'}(\omega,T) =  \delta_{\sigma \sigma'} \delta_{\alpha \alpha'}
-S_{\sigma \alpha  , \sigma' \alpha'} \mathcal{G} \left(
  \frac{\omega}{T^*},\frac{T}{T^*} \right)~
\end{eqnarray}
where $S_{\sigma \alpha  , \sigma' \alpha'}$ is the scattering
S matrix, which is an $\omega=0$ and $T=0$ quantity
characterizing the FL FP. For the 2CK model it is given by,
\begin{equation}
\label{s2ck}
S^{2CK}_{\sigma \alpha, \sigma' \alpha'}=\left [- \delta_{\sigma
  \sigma'}(\vec{\lambda}_f \cdot \vec{\tau}_{\alpha \alpha'})+ i
(\vec{\lambda}_B \cdot \vec{\sigma}_{\sigma \sigma'}) \delta_{\alpha
  \alpha'}\right ]/\lambda,
\end{equation}
with $\vec{\lambda}_f=\{ \lambda_2,\lambda_3,\lambda_1\}$. For the 2IK model it is,
\begin{equation}
\begin{split}
\label{s2ik}
S^{2IK}_{\sigma \alpha, \sigma' \alpha'} = &\Big [-\lambda_1 \delta_{\sigma \sigma'} \delta_{\alpha \alpha'}
+ i \delta_{\sigma \sigma'} (\lambda_2 \tau^x_{\alpha \alpha'}
+\lambda_3 \tau^y_{\alpha \alpha'}) \\
& + i (\vec{\lambda}_B  \cdot \vec{\sigma}_{\sigma \sigma'}) \tau^z_{\alpha \alpha'} \Big ]/\lambda.
\end{split}
\end{equation}

The single function $\mathcal{G} $
describes the crossover due
to a generic combination of relevant perturbations in both 2CK and
2IK models. It does not depend on details of the model or the
particular perturbations present, except through the resulting crossover
scale $T^*$. Thus $ \mathcal{G} ( \tilde{\omega} ,\tilde{T} )$
is a \emph{universal} function of rescaled energy
$\tilde{\omega}=\omega/T^*$ and temperature $\tilde{T}=T/T^*$.
Our exact result at finite temperature is,
\begin{equation}
\label{analyticformula}
\begin{split}
&\mathcal{G} \left(\tilde{\omega},\tilde{T} \right)=\frac{ \frac{-i}{\sqrt{2 \pi^3 \tilde{T}}}}{\tanh \frac{\tilde{\omega}}{2 \tilde{T}}} \frac{\Gamma \left(\frac{1}{2}+\frac{1}{2 \pi \tilde{T}} \right)}{\Gamma \left(1+\frac{1}{2 \pi \tilde{T}} \right)}\times  \\
&\int_{-\infty}^\infty dx \frac{ e^{ \frac{ix \tilde{\omega}}{\pi \tilde{T}}}}{\sinh x} {\rm{Re}} \left[ _{2}F_1\left(\frac{1}{2},\frac{1}{2};1+\frac{1}{2 \pi \tilde{T}},\frac{1-\coth x}{2} \right) \right],
\end{split}
\end{equation}
where $\Gamma$ is the Gamma function, and $_{2}F_1(a,b,c,z) $ is the
Gauss hypergeometric function.\cite{Abramowitz} At $T=0$,
Eq.~\ref{analyticformula} reduces\cite{note:convention} to the result of
Ref.~\onlinecite{SelaMitchellFritz},
\begin{equation}
\label{exactT=0}
\mathcal{G}(\tilde{\omega},0) = \frac{2}{\pi} K\left[-i \tilde{\omega}  \right],
\end{equation}
where $K[z]$ is the complete elliptic integral of the first
kind, yielding asymptotically $\mathcal{G}(\tilde{\omega},0) = 1- i \tilde{\omega}/4-(3\tilde{\omega}/8)^2+\mathcal{O}(\tilde{\omega}^3)$ for $\tilde{\omega} \ll 1$; and $\mathcal{G}(\tilde{\omega},0) =
\frac{\sqrt{i}}{2\pi}(\pi-2i\log[16 \tilde{\omega}]
)\tilde{\omega}^{-1/2} +\mathcal{O}(\tilde{\omega}^0)$
for $\tilde{\omega} \gg 1$.

Below we consider the local density of states (spectrum)
$t_{\alpha\sigma}(\omega,T)$, from which conductance can be calculated
(see Eqs.~\ref{zbcond} and \ref{asymcond}). It is related to the t
matrix via Eq.~\ref{specdef}, and is thus given exactly along
the NFL to FL crossover by
Eqs.~\ref{exactresult}--\ref{analyticformula}:
\begin{equation}
\label{specexact}
t_{\sigma\alpha}(\omega,T)=\tfrac{1}{2}-\tfrac{1}{2}\text{Re}\left [ S_{\sigma\alpha,\sigma\alpha}\mathcal{G} \left(\tilde{\omega},\tilde{T} \right)\right ],
\end{equation}
where the required diagonal elements of the full S matrix
(Eqs.~\ref{s2ck} and \ref{s2ik}) are more simply expressed as,
\begin{equation}
\label{Smatrix}
S^{2CK}_{\sigma \alpha,\sigma \alpha} = (- \alpha \lambda_1+i\sigma
  \lambda_B^z)/\lambda= \alpha S^{2IK}_{\sigma \alpha,\sigma \alpha}
\end{equation}
with $\sigma=\pm 1$ for spins $\uparrow/\downarrow$ and $\alpha=\pm 1$
for channel $L/R$ (and we use $\vec{\lambda}_B \parallel \hat{z}$ for
simplicity). For $\lambda_f^x=\lambda_f^y=0$ and $\lambda_B^x=\lambda_B^y=0$ scattering preserves channel and spin, and  the FL phase shift $\delta_{\sigma\alpha}$ then follows
from $S_{\sigma\alpha,\sigma \alpha}=\exp[2i\delta_{\sigma\alpha}]$.

These exact results for the crossover are compared with
finite-temperature NRG calculations in Sec.~\ref{nrgcomp}, with
excellent agreement.  

We now examine the generic behavior of the spectral function at
finite temperatures in the crossover regime.
Although we consider explicitly $L$-channel spectra $t_{\sigma
  L}(\omega,T)$ in the following, note from Eq.~\ref{Smatrix} that $t_{\sigma
  L}(\omega,T)\leftrightarrow t_{\sigma R}(\omega,T)$ upon
exchanging $\Delta_z \leftrightarrow -\Delta_z$ in the 2CK model, or
$\vec{B}_s\leftrightarrow -\vec{B}_s$ in the 2IK model. Also,
$t_{\uparrow \alpha}(\omega,T) \leftrightarrow t_{\downarrow
  \alpha}(\omega,T)$ on reversing the magnetic field,
$\vec{B}\leftrightarrow -\vec{B}$ (and in the zero-field case,
$\sigma=\uparrow$ and $\downarrow$ spectra are of course identical).

\begin{figure}[t]
\begin{center}
\includegraphics*[width=80mm]{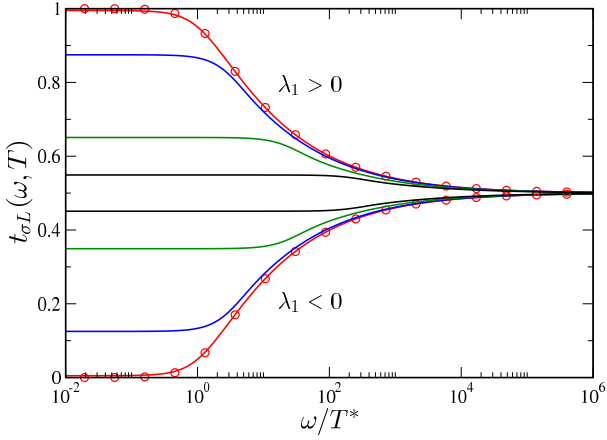}
\caption{\label{fig:k} Spectrum $t_{\sigma L}(\omega,T)$ vs $\omega/T^*$ for
$T/T^*=10^{-1}, 1, 10, 10^2$, approaching $t_{\sigma L}=\tfrac{1}{2}$
from above ($\lambda_1>0$) or
below ($\lambda_1<0$). Circles show $T=0$ result of Eq.~\ref{exactT=0}.
}
\end{center}
\end{figure}

In Fig.~\ref{fig:k}, we take the representative case of finite channel
anisotropy $\Delta_z$ in the 2CK model, or finite detuning $(K-K_c)$
in the 2IK model, and plot $t_{\sigma L}(\omega,T)$
as a full function of $\omega/T^*$ for different temperatures
$T/T^*$. Since only $\lambda_1$ acts in either case,
$S_{\sigma\alpha,\sigma'
  \alpha'}=\pm\delta_{\sigma\sigma'}\delta_{\alpha\alpha'}$ is diagonal (see
Eqs.~\ref{s2ck}, \ref{s2ik}), meaning that an electron in channel
$\alpha$ scatters elastically at low energies, and stays in channel
$\alpha$. By Eq.~(\ref{specexact}), the spectrum $t_{\sigma
  \alpha}(\omega,T)$ then probes the
real part of the universal function $\mathcal{G}$ because
$S_{\sigma\alpha,\sigma \alpha}$ is real.

General scaling arguments suggest that RG flow stops on an energy scale
given by the temperature. As seen from Fig.~\ref{fig:k}, this is
indeed the case, with the spectrum $t_{\sigma L}(\omega,T)\simeq
t_{\sigma L}(0,T)$  essentially constant for $|\omega| \ll T$.
Mutatis mutandis, for $T \ll T^*$ one obtains $t_{\sigma
  L}(\omega,T)\simeq t_{\sigma L}(\omega,0)$, corresponding to the
$T=0$ limit considered previously.\cite{SelaMitchellFritz}
At $T=0$ and $\omega=0$, Eq.~\ref{specexact} yields
$t_{\sigma\alpha}(0,0)=\tfrac{1}{2}-\tfrac{1}{2}\text{Re}S_{\sigma\alpha,\sigma\alpha}$,
which is determined solely by the S matrix and hence the phase shift
associated with the stable FL FP. When only $\lambda_1$ acts,
the spectrum is thus $t_{\sigma \alpha}(0,0)=0$ or $1$ only (with
corresponding phase shifts $0$ or $\pi/2$). In particular, the
Kondo phase is characterized by unitarity $t_{\sigma \alpha}(0,0)=1$,
obtained in the more strongly coupled channel for the 2CK
model, and in both channels for $K<K_c$ in the 2IK model.

In the opposite limit $T\gg T^*$ ($\ll T_K$) RG flow to the FL FP is
completely cut off, and inelastic scattering\cite{borda} at the NFL FP
results in $t_{\sigma\alpha}(\omega,T)\simeq \tfrac{1}{2}$ for all
$|\omega|\ll T_K$.

\begin{figure}[t]
\begin{center}
\includegraphics*[width=80mm]{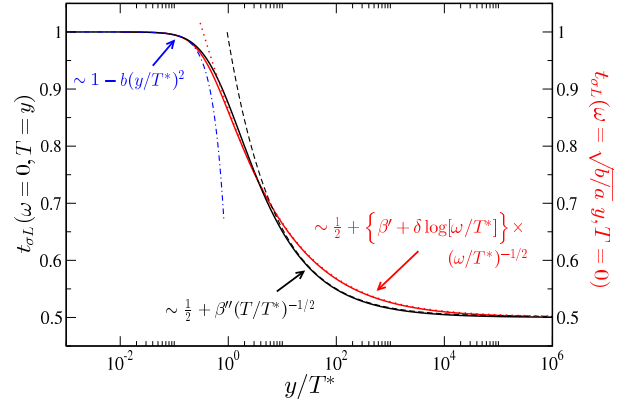}
\caption{\label{fig:tw} $t_{\sigma L}(\omega=0,T=y)$ and $t_{\sigma
    L}(\omega=y\sqrt{b/a},T=0)$ vs $y/T^*$ for
  $\lambda_1>0$. Common FL asymptote Eq.~\ref{FLspec} shown as dot-dashed
  line; NFL asymptotes Eq.~\ref{NFLspec} shown as dashed and dotted lines.
}
\end{center}
\end{figure}

The generic RG picture illustrated in Fig.~\ref{fig:pd} and supported
by Fig.~\ref{fig:k}, suggests an approximate complementarity between
$\omega$ and $T$.
This is explored further in Fig.~\ref{fig:tw}, where we compare the
zero-frequency value of the spectrum as a function of temperature, with
the zero-temperature spectrum as a function of frequency.
As immediately seen, there is a striking similarity.
Indeed, a classic signature of FL physics (arising at
low-energies/temperatures $|\omega|,T \ll T^*$)
is the common quadratic dependence of the spectrum
on both frequency and temperature,
\begin{equation}
\label{FLspec}
t_{\sigma\alpha}(\omega,T) ~~\overset{FL}{\sim} ~~ t_{\sigma\alpha}(0,0) +a \bigg (
  \frac{\omega}{T^*} \bigg )^2 + b \bigg (\frac{T}{T^*} \bigg )^2,
\end{equation}
with $a$ and $b$ constants $\mathcal{O}(1)$ that depend on the
particular model under consideration. Perturbation
theory with respect to the FL FP in the spirit of
Nozi\`{e}res,\cite{Nozieres} yields $a/b=9/(7\pi^2)$
for the 2IK model (see Appendix~\ref{pertFL}). This same ratio also
follows directly from the limiting behavior of the full crossover function,
Eq.~\ref{analyticformula}, which yields $|a|=9/128\approx 0.07$ and
$|b|=7\pi^2/128\approx 0.54$; and as such provides a stringent check
of our results.

But the exact symmetry between $\omega$ and $T$ in Eq.~\ref{FLspec} is
a special property of the FL FP itself, and does not in general apply
at higher energies; although as seen from Fig.~\ref{fig:tw}, the
qualitative behavior over the full crossover is in fact rather
similar. In the vicinity of the NFL FP (arising for $T^*\ll
\max(\omega, T) \ll T_K$), Eq.~\ref{analyticformula} gives
asymptotically,
\begin{subequations}
\label{NFLspec}
\begin{align}
\label{asymptoticT}
t_{\sigma L}(\omega,T=0) &\overset{NFL}{\sim} \tfrac{1}{2}
\pm\left \{\beta' + \delta\log  \frac{\omega}{T^*} \right \} \left( \frac{\omega}{T^*}\right)^{-\tfrac{1}{2}}\\
\label{asymptoticw}
t_{\sigma L}(\omega=0,T) &\overset{NFL}{\sim} \tfrac{1}{2}
\pm \beta'' \left( \frac{T}{T^*}\right)^{-\tfrac{1}{2}},
\end{align}
\end{subequations}
with $\pm$ for $\lambda_1\gtrless 0$; and $\beta' = \frac{(1+\pi) (2 \gamma -\pi + \log 64)}{2^{5/2} \pi} \approx 0.5061$ (where $\gamma$ is Euler's constant),
$\delta =-\frac{1}{2^{3/2} \pi} \approx -0.1125$, and
$\beta''\approx 0.4925$. Terms of the form $(\omega/T^*)^{-1/2}$ and
$(T/T^*)^{-1/2}$ in Eq.~\ref{NFLspec} signal the scaling dimension
$1/2$ of the relevant perturbation. Whereas such powerlaws occur in
both the frequency and temperature dependence,  
additional logarithmic corrections appear in the
frequency-dependence only. This difference can be understood by comparing
the full $T=0$ result (Eq.~\ref{exactT=0}) with the high-$T$ behavior
captured by perturbation theory around the NFL FP (see
Ref.~\onlinecite{ising_note} and Appendix~\ref{NFLpert}).
The full dependence on $\omega$ and $T$ described by
Eq.~\ref{analyticformula} naturally leads to more subtle behavior when
$|\omega|$ and $T$ are of the same order, as shown in Fig.~\ref{fig:k}.

When some degree of inter-channel charge transfer is also present, the
NFL to FL crossover is generated by the \emph{combination} of relevant
perturbations $\lambda_1$ and $\lambda_2$. In the 2CK model, this
corresponds to finite channel anisotropy $\Delta_z$ and
impurity-mediated tunneling $\Delta_x$ (see Eq.~\ref{delta2CK}); while
for 2IK it corresponds to finite detuning $(K-K_c)$
and direct tunneling $V_{LR}$ (see Eq.~\ref{delta2IK}).
The resulting behavior in the 2CK model can be simply understood
because the perturbations $\Delta_z$ and $\Delta_x$ are related by a `flavor'
rotation of the bare Hamiltonian, as discussed further in
Sec.~\ref{flavorrot}. The 2IK model does not possess such a
flavor symmetry, although an \emph{emergent}
symmetry\cite{CFT2CK,CFT2IKM} of the NFL FP Hamiltonian
can be exploited when $T^*\ll T_K$.
In fact, as shown in Sec.~\ref{unitarytrans}, this symmetry allows all
the relevant perturbations in either 2CK or 2IK models to be simply
related,\cite{SelaMitchellFritz} implying the
existence of a \emph{single} crossover function.

The rotation $(\lambda_1,\lambda_2)\rightarrow \tilde{\lambda}_1$ can
be used to relate systems where both $\lambda_1$ and $\lambda_2$ act,
to those in which $\lambda_1$ alone acts (as in Fig.~\ref{fig:k}).
But $t_{\sigma\alpha}(\omega,T)$ probes the system in the
original unrotated basis, and hence the spectra undergo a
rescaling when finite $\lambda_2$ is included:
$t_{\sigma\alpha}(\omega,T)\rightarrow \tfrac{1}{2}+\left |
  \tfrac{\lambda_1}{\lambda}\right
|[\tilde{t}_{\sigma\alpha}(\omega,T)-\tfrac{1}{2}]$.
In particular, the spectral function is totally flattened the limit
$|\lambda_2|\gg |\lambda_1|$, with $t_{\sigma\alpha}\simeq
\tfrac{1}{2}$ at both FL and NFL FPs. At the FL FP,
electrons thus scatter elastically between $\alpha=L$ and $R$
channels; and the corresponding S matrix $|S_{\sigma\alpha,\sigma'
  \alpha'}|=\delta_{\sigma\sigma'}(1-\delta_{\alpha\alpha'})$ is
purely \emph{off}-diagonal when inter-channel charge
transfer dominates (see Eqs.~\ref{s2ck}, \ref{s2ik}).
Thus, no crossover shows up in the spectrum or conductance, although
$T^*$ is of course finite (see Eq.~\ref{tstar}); and the crossover can
still appear in other physical quantities.\cite{akm:oddimp}

\begin{figure}[t]
\begin{center}
\includegraphics*[width=80mm]{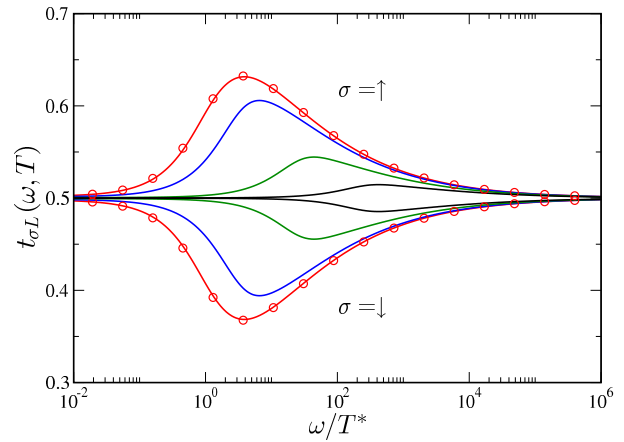}
\caption{\label{fig:mag} $t_{\sigma L}(\omega,T)$ vs $\omega/T^*$ for
$T/T^*=10^{-1}, 1, 10, 10^2$, in the presence of finite uniform
(staggered) magnetic field in the 2CK (2IK) model,
$\lambda_B^z>0$. Spectra approach $t_{\sigma L}=\tfrac{1}{2}$ from
above ($\sigma=\uparrow$) or below ($\sigma=\downarrow$).
Circles show $T=0$ result.
}
\end{center}
\end{figure}

When a uniform (2CK) or staggered (2IK) magnetic field acts (finite
$\lambda_B^z$ only), $S_{\sigma\alpha,\sigma
  \alpha}=\pm i$ is pure imaginary (with phase shifts
$\delta_{\sigma\alpha}=\pm\pi/4$), and again we obtain
$t_{\sigma\alpha}(0,0)=\tfrac{1}{2}$ at the FL FP. However,
$t_{\sigma\alpha}(\omega,T)$ now probes the imaginary part of the
universal function $\mathcal{G}$ (see Eq.~\ref{specexact}); and so the full spectrum along the NFL to FL
crossover due to $\lambda_B^z$ is simply the Hilbert transform of the
spectrum due to $\lambda_1$ --- compare Figs.~\ref{fig:k} and
\ref{fig:mag}.

A spectral feature in consequence appears on the intermediate scale of
$T^*$ for finite $\lambda_B^z$, even though
$t_{\sigma\alpha}=\tfrac{1}{2}$ at both NFL and FL FPs; as shown in
Fig.~\ref{fig:mag}. The existence of such a feature can be understood
physically from the impurity magnetization $M\sim B^z$ arising for
small applied field $B^z$ in the 2CK model (or staggered magnetization
$M_s\sim B_s^z$ due to a staggered field in the 2IK
model). Since the magnetization $M(T)\propto \int_{-\infty}^{\infty}
d\omega~f(\omega,T) [ t_{\uparrow\alpha}(\omega,T)
-t_{\downarrow\alpha}(\omega,T) ]\ne 0$ is finite for finite applied field,
$t_{\uparrow\alpha}(\omega,T) \ne t_{\downarrow\alpha}(\omega,T)$.
A `pocket' thus opens between the `up' and `down' spin spectra at
$|\omega|\sim T^*$, whose area is proportional to the magnetization.
In particular, the temperature-dependence of the magnetization can be
extracted from the universal function, Eq.~\ref{analyticformula}, viz
\begin{equation}
\label{magnetization}
 M(T)\propto \int_{-T_K}^{T_K}
 d\omega~f(\omega,T)~\text{Im}\mathcal{G}\left (
   \frac{\omega}{T^*},\frac{T}{T^*}\right ),
\end{equation}
valid for small perturbations $\lambda_B^z$, such that $T^*\ll T_K$ as
usual (the high-frequency cutoff $|\omega|\sim T_K$ then being
justified since $t_{\uparrow\alpha}(\omega,T) \simeq t_{\downarrow\alpha}(\omega,T)$
for $|\omega|\gtrsim T_K$, as confirmed directly from NRG).

\begin{figure}[t]
\begin{center}
\includegraphics*[width=70mm]{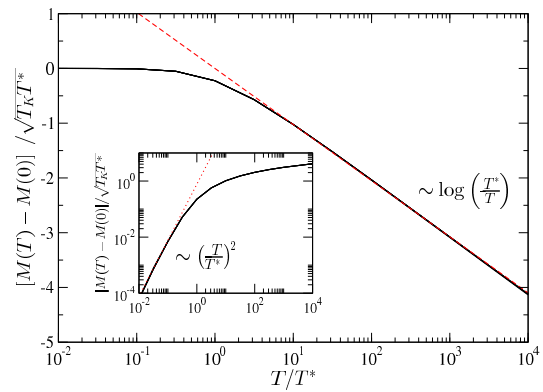}
\caption{\label{fig:magn} Magnetization $[M(T)-M(0)]/\sqrt{T_KT^*}$ vs
$T/T^*$ for $\lambda_B^z>0$ and $T_K/T^*=10^6, 10^7, 10^8, 10^9$. NFL
asymptote Eq.~\ref{magNFL} shown as
dashed line; FL asymptote Eq.~\ref{magFL} shown as dotted line in the
inset.
}
\end{center}
\end{figure}

Using Eq.~\ref{exactT=0}, the zero-temperature magnetization depends
on $T_K$ via,
\begin{equation}
\label{m0}
 M(0)=c\sqrt{T_KT^*}\log \left (\frac{T_K}{T^*} \right ),
\end{equation}
and since $T^*\sim (B^z)^2/T_K$ (see Eq.~\ref{tstar}), it follows that
$M(0)\sim B^z\log(T_K/B^z)$. The full temperature dependence $M(T)$ vs
$T$ is shown in Fig.~\ref{fig:magn}, demonstrating scaling collapse
for different Kondo scales $T_K$. The asymptotic behavior in the
vicinity of the FL and NFL FPs follows as,
\begin{subequations}
\begin{align}
\label{magFL}
\frac{1}{\sqrt{T_KT^*}} \left [M(T)-M(0) \right ]~~~
&\overset{T/T^*\ll 1}{\sim}~~~ -d\left ( \frac{T}{T^*} \right )^2,\\
\label{magNFL}
&\overset{T/T^*\gg 1}{\sim}~~~ -c\log \left (\frac{T}{T^*} \right ),
\end{align}
\end{subequations}
yielding in particular $M(T)\sim B^z\log(T_K/T)$ when $T^*\ll T \ll
T_K$. This asymptotic behavior can again be understood from
perturbation theory around the FL and NFL FPs. Furthermore, since
$\chi_{\text{imp}}(T) = \lim_{B^z\rightarrow 
  0} M(T)/B^z$, when $T^*\ll B^z\ll T \ll T_K$, one obtains
\begin{equation}
\label{chimag}
\chi_{\text{imp}}(T) \sim \log(T_K/T),
\end{equation}
for the uniform (staggered) magnetic susceptibility of the 2CK (2IK)
model. This diverging susceptibility is a classic signature of the NFL
FP, known for example from the Bethe  ansatz solution of the 2CK
model,\cite{andrei} or from CFT for the 2IK model.\cite{CFT2IKM}
However, at lower temperatures $T\ll B^z$, $M(T)/B^z$ does not
correspond to the magnetic susceptibility: here the NFL to FL
crossover itself is being probed. Indeed, for
$T\ll T^*\ll B^z$, we obtain a quadratic $(T/T^*)^2$
temperature-dependence of magnetization, Eq.~\ref{magFL},
characteristic of the FL FP.

\begin{figure}[t]
\begin{center}
\includegraphics*[width=75mm]{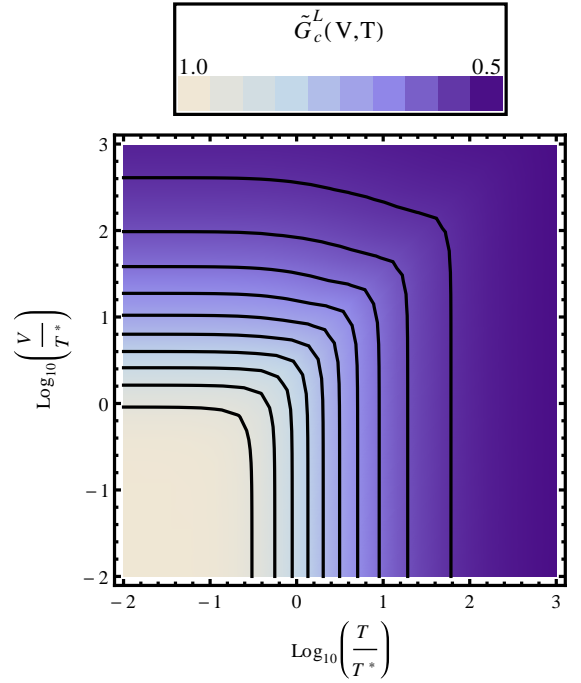}
\caption{\label{fig:condcp} Conductance
  $\tilde{G}_c^{L}(V,T)\equiv G_c^{L}(V,T)/(2e^2 h^{-1}
  G_0^{L})$ vs bias $V$ and temperature $T$ for
  $\lambda_1>0$. Black lines connect regions of constant
  conductance. Light colors correspond to high conductance near the FL
  FP; dark colors correspond to lower conductance near the NFL FP.
}
\end{center}
\end{figure}

\begin{figure}[t]
\begin{center}
\includegraphics*[width=75mm]{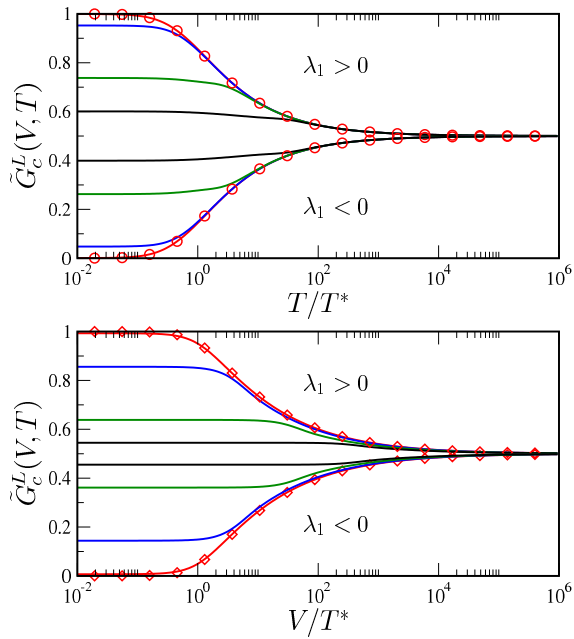}
\caption{\label{fig:condcut} Conductance
  $\tilde{G}_c^{L}(V,T)\equiv G_c^{L}(V,T)/(2e^2 h^{-1}
  G_0^{L})$ arising for finite $\lambda_1$.
\emph{Upper panel:} vs temperature $T/T^*$ for $V/T^*=10^{-1},
1, 10, 10^2$, approaching $\tilde{G}_c^{L}=\tfrac{1}{2}$ from above
($\lambda_1>0$) or  below ($\lambda_1<0$). Circles show the exact
zero-bias result. \emph{Lower panel:} vs bias $V/T^*$ for
$T/T^*=10^{-1}, 1, 10, 10^2$. Diamonds show zero-temperature result.}
\end{center}
\end{figure}

Finally, we turn to conductance $\tilde{G}_c^{\alpha}(V,T)$, obtained
from the spectrum $t_{\sigma\alpha}(\omega,T)$ by combining
Eqs.~\ref{asymcond}, \ref{analyticformula} and \ref{specexact}. It
follows as
\begin{equation}
\label{condexact}
\begin{split}
&\tilde{G}_c^{\alpha}(V,T)=\frac{1}{2}-\frac{\Gamma(\tfrac{1}{2} +
  \tfrac{1}{2\pi\tilde{T}})}{(8\pi \tilde{T})^{3/2}\Gamma(1+
\tfrac{1}{2\pi\tilde{T}})}\sum_{\sigma}\text{Im} \Bigg \{
S_{\sigma\alpha,\sigma\alpha} \times \\
&\int_{-\infty}^{\infty} dx ~I(V,T,x) {\rm{Re}}
\left[ _{2}F_1\left(\frac{1}{2},\frac{1}{2};1+\frac{1}{2 \pi
      \tilde{T}},\frac{1-\coth x}{2} \right) \right]  \Bigg \},
\end{split}
\end{equation}
where the integral over $\omega$ can be evaluated using contour
methods,
\begin{equation}
\label{wint}
\begin{split}
I(V,T,x)=&\int_{-\infty}^{\infty}d\tilde{\omega}~\frac{\exp\left ({\frac{i x
      \tilde{\omega}}{\pi\tilde{T}}}\right )\text{sech}^2\left(\frac{\tilde{\omega}-\tilde{V}}{2\tilde{T}}\right)}{\sinh(x)\tanh\left(\frac{\tilde{\omega}}{2\tilde{T}}\right)}\\
=&2\pi i \tilde{T}~
\text{csch}^2(x)~\text{sech}^2\left(\tfrac{\tilde{V}}{2\tilde{T}}\right) \Bigg [
  \cosh(x)\\
&~~~-\exp\left(\tfrac{i x \tilde{V}}{\pi\tilde{T}}\right )\left ( 1+\tfrac{i
      x}{\pi}\sinh\left (\tfrac{\tilde{V}}{\tilde{T}}\right)\right ) \Bigg ],
\end{split}
\end{equation}
with rescaled $\tilde{V}=V/T^*$, $\tilde{T}=T/{T^*}$,
$\tilde{\omega}=\omega/T^*$ as before. In particular at zero-bias,
\begin{equation}
\label{wintv0}
I(V=0,T,x)=\frac{\pi i \tilde{T}}{\cosh^2\left(\tfrac{x}{2}\right)}.
\end{equation}

A color plot of conductance $\tilde{G}_c^{\alpha}(V,T)$ along the NFL
to FL crossover is shown in Fig.~\ref{fig:condcp}, for the
representative case of $\lambda_1>0$. The black lines connect regions
of equal conductance. In the FL regime $V,T\ll T^*$, the quadratic
form of Eq.~\ref{FLspec} thus yields a simple ellipse; while in the NFL
regime there is a pronounced $V$-$T$ anisotropy. The detailed behavior
of Eq.~\ref{condexact} is seen on taking cuts through Fig.~\ref{fig:condcp} at
constant $V$ and $T$, as shown in Fig.~\ref{fig:condcut}.

Returning to the 2CK quantum dot system of
Ref.~\onlinecite{Potok07}, we comment now on the possible strength of
symmetry-breaking perturbations present in the experiment. 
Due to the Coulomb blockade physics of the quantum box, inter-channel
charge-transfer was effectively suppressed. In the absence of a
magnetic field, the dominant perturbation is thus
\emph{channel anisotropy}, $\Delta_z$ (see Eq.~\ref{delta2CK}). The experiment
showed\cite{Potok07} 2CK scaling of conductance around
$T_K$, but no FL crossover at $T^*$. This implies $T^*\ll T \ll
T_K$ --- see for example Fig.~\ref{fig:k} for $T/T^*=100$, which shows
little sign of the crossover. 
Since $T/T_K\approx 0.1$ in the experiment and taking $T/T^*>100$,
from Eq.~\ref{tstar} and Table~\ref{table:pert} it follows that $c_1
\nu \Delta_z $ could be at most $\sim 0.03$ ($c_1=\mathcal{O}(1)$ depends
on details of the model/device setup). The observed 2CK 
physics is thus an impressive testament to the tunability and control
available in such quantum dot devices. 

Having presented our main results and discussed their physical implications, we
turn in the following sections to the formal derivation.


\section{Exact $T=0$ crossover Green function in the 2IK model}
\label{2ikgf}

Our goal is an exact expression for the NFL to FL crossover t matrix,
which is related to the electron Green function.
In Ref.~\onlinecite{SelaMitchellFritz}, we calculated the crossover at
$T=0$; further details of that calculation are presented here,
providing as they do the necessary foundations for our generalization
of the results to finite temperature.

\subsection{Fixed point Hamiltonians and Green functions}
\label{chiralG}

The structure of the NFL fixed point Hamiltonian of the 2IK model
allows for an elegant 
description of the NFL to FL crossover.\cite{SelaMitchellFritz} Before
presenting that derivation, we discuss first some relevant
preliminaries which will be of later use. In particular, we consider
now the representation of the fixed point Hamiltonians within CFT,
and the structure of the corresponding Green functions.

Our starting point is a description of the free conduction electron
Hamiltonian $H_0$ in terms of chiral Dirac fermions.
A 1D quadratic dispersion relation
$\epsilon(\vec{k})\equiv \epsilon_k = k^2/2m -\epsilon_F$ can be
linearized near the Fermi points $k=\pm k_F$, $\epsilon_k \simeq \pm v_F(k \mp k_F)$ (with
$v_F$ the Fermi velocity). This is the standard
case\cite{hewson} and applies to arbitrary dimension within the assumption that the bare density of states is flat at
low energies.\cite{affleckreview} Conduction
electron operators can be Fourier transformed and expanded near the
Fermi points, focusing on states within width $2D \ll v_F k_F$ around $\epsilon_F$,
\begin{eqnarray}
\Psi_{\sigma \alpha}(x) &=& \sum_k e^{i k x} \psi_{k \sigma \alpha}  \\ & \cong &e^{i k_F x} \sum_{k = k_F-D/v_F}^{k_F+D/v_F}  \psi_{k \sigma \alpha} e^{i (k-k_F)x} + (k_F \to - k_F) \nonumber.
\end{eqnarray}
We thus define left and right movers,\cite{affleckreview}
\begin{equation}
\label{chiralft}
\psi_{(l,r)\sigma\alpha}(x)=\sum_{k=-D/v_F}^{D/v_F} ~e^{ i k x}\psi_{k\mp k_F,\sigma \alpha},
\end{equation}
defined for $x\ge 0$, with $x$ the distance from the impurities
located at the `boundary' $x=0$. With a boundary condition
$\psi_{r,\sigma\alpha}(0)=\psi_{l,\sigma\alpha}(0)$ we introduce  a
single left-moving chiral Dirac fermion for all $x$,
$\psi_{\sigma\alpha}(x)=\theta(x)\psi_{l,\sigma\alpha}(x)+\theta(-x)\psi_{r,\sigma\alpha}(-x)$. Since
the new operators satisfy the usual fermionic 
anticommutation relations $\{
\psi^{\dagger}_{\sigma\alpha} (x_1) ,
\psi^{\phantom{\dagger}}_{\sigma'\alpha'} (x_2) \} =
\delta_{\sigma\sigma'}\delta_{\alpha\alpha'}\delta(x_1-x_2)$,
the free Hamiltonian
$H_0$ can be expressed as a chiral Hamiltonian,\cite{affleckreview}
 \begin{equation}
\label{h0}
 H_0=v_F\sum_{\sigma,\alpha}\int_{-\infty}^\infty \textit{dx}~ \psi^{\dagger}_{\sigma\alpha} i \partial_x \psi^{\phantom{\dagger}}_{\sigma\alpha}.
 \end{equation}
Hereafter, we set $v_F\equiv 1$ [and the Fermi level density of
states is then $\nu=1/(2\pi v_F) \equiv 1/(2 \pi)$].

The free electron Green function then follows as,\cite{bosonizationforbeginners}
\begin{equation}
\begin{split}
\label{freeG}
\langle
\psi_{\sigma\alpha}^{\phantom{\dagger}}(\tau,x_1)\psi_{\sigma'\alpha'}^{\dagger}(0,x_2)\rangle_0=&\frac{\frac{1}{2\pi}\frac{\pi}{\beta}\delta_{\sigma \sigma'}  \delta_{\alpha \alpha'}}{\sin[\frac{\pi}{\beta}
  (\tau + i x_1-i x_2)]} \\
\equiv& \delta_{\sigma \sigma'}  \delta_{\alpha \alpha'}G^0(\tau,x_1-x_2),
\end{split}
\end{equation}
where $\tau$ is imaginary time. The corresponding Matsubara Green
function is then defined by the transformations
\begin{equation}
\label{G0matsubara}
\begin{split}
&G^0(\tau,x_1-x_2) = \frac{1}{\beta}
\sum_n e^{- i \omega_n \tau}  \mathcal{G}^0(x_1-x_2,i \omega_n) \\
&\mathcal{G}^0(x_1-x_2,i \omega_n) =
\int_{-\beta/2}^{\beta/2} \textit{d}\tau ~e^{i \omega_n \tau}  G^0(\tau,x_1-x_2),
\end{split}
\end{equation}
with Matsubara frequencies $\omega_n=\tfrac{\pi}{\beta}(1+2n)$ for
integer $n$, and where $\beta=1/k_B T$ is inverse temperature.
Direct evaluation of Eq.~\ref{G0matsubara} using Eq.~\ref{freeG} then
yields simply,
\begin{equation}
\begin{split}
\label{G0}
&\mathcal{G}^0(x_1-x_2,i \omega_n) =i e^{ \omega_n ( x_1- x_2)} \times\\
& \delta_{\sigma \sigma'}  \delta_{\alpha \alpha'}[\theta( \omega_n) \theta(x_2-x_1)-\theta(- \omega_n) \theta(x_1-x_2)].
\end{split}
\end{equation}

Of course, the interesting behavior arises when coupling to the
impurities is switched on. As usual,\cite{hewson} the full Green
function is related to the  t matrix via
\begin{equation}
\label{Tdefinition}
\begin{split}
  \boldsymbol{\mathcal{G}}(x_1,x_2,i \omega_n)
  =\boldsymbol{\mathcal{G}}^0(x_1-x_2,i \omega_n) \\
+\boldsymbol{\mathcal{G}}^0(x_1,i \omega_n) \boldsymbol{\mathcal{T}}(i \omega_n) \boldsymbol{\mathcal{G}}^0(-x_2,i \omega_n),
\end{split}
\end{equation}
where $\boldsymbol{\mathcal{G}}$,
$\boldsymbol{\mathcal{G}}^0$ and
$\boldsymbol{\mathcal{T}}$ are $4 \times 4$ matrices
with indices taking the values $\sigma\alpha =\uparrow L, \downarrow
L, \uparrow R, \downarrow R$. In particular, it should be
noted that the t matrix is local. Eqs.~\ref{G0} and \ref{Tdefinition} also
imply that if $x_1$ and $x_2$ have equal sign then the full Green
function reduces to the free Green function, reflecting the chiral
nature of the Dirac fermion, Eq.~\ref{h0}. Since all such fermions are
now left-moving, information about scattering from the impurities
located at the boundary $x=0$ --- and hence the t matrix --- is
obtained from the full Green function with $x_1$ and $x_2$ located on
\emph{opposite} sides of the boundary.

The free Green function Eq.~\ref{freeG} is naturally obtained at FPs where the free boundary condition pertains. 
But at any conformally invariant FP, the powerful machinery of boundary CFT gives nonperturbative information about the Green
function. Specifically, when the boundary condition is obtained from fusing with some primary field $a$, then correlation functions are given generically by\cite{Cardy91}
 \begin{equation}
 \label{corrfunc}
 \langle \mathcal{O}^{\phantom{\dagger}}_d(\tau,i x_1)\mathcal{O}^{\dagger}_d(0,i x_2) \rangle = \frac{\frac{S_a^d /S_0^d}{S_a^0 /S_0^0}}{[\tau+ i x_1 - ix_2]^{2 d}},
 \end{equation} 
 where $d$ is the scaling dimension of the primary field
 $\mathcal{O}$, and $S_j^a$ are elements of the modular S
 matrix.\cite{Cardy91} 
 Using the conformal mapping from the plane to the cylinder with
 circumference $\beta$, one obtains\cite{Cardy91} the generalization
 to finite temperature $T=\beta^{-1}$, 
 \begin{equation}
 \label{corrfuncT}
 \langle \mathcal{O}^{\phantom{\dagger}}_d(\tau,i x_1)\mathcal{O}^{\dagger}_d(0,i x_2) \rangle = \frac{\frac{S_a^d /S_0^d}{S_a^0 /S_0^0}}{(\tfrac{\beta}{\pi}\sin [\tfrac{\pi}{\beta}(\tau+ i x_1 - i x_2)])^{2 d}}.
 \end{equation} 
 
In the present context, we are interested in the electron Green function at the conformally invariant free fermion, NFL and FL FPs of the 2IK model. With $\mathcal{O}=\psi_{\sigma\alpha}$ the $d=1/2$ fermion field, the full FP Green functions takes the form,\cite{CFT2IKM}
\begin{equation}
\label{cibc}
 G_{\sigma\alpha,\sigma'\alpha'}^{BCFT}(\tau,x_1-x_2) =\frac{\frac{1}{2\pi} \frac{\pi}{\beta} S_{\sigma \alpha,\sigma' \alpha'}}{\sin[\frac{\pi}{\beta} (\tau + i x_1-i x_2)]},
 \end{equation}
where $\boldsymbol{S}$ can be understood as the one-particle to one-particle scattering
matrix, and can be calculated from the modular S matrix in
the case of boundary conditions obtained by fusion.\cite{Cardy91}  
Thus, the effective FP theory is identical to that of the free
theory discussed above, but with a modified boundary condition that determines the scattering matrix $\boldsymbol{S}$.

Choosing $x_1>0$ and $x_2<0$ in Eq.~\ref{Tdefinition} yields
$i\boldsymbol{\mathcal{T}} = \boldsymbol{\mathds{1}}- \boldsymbol{S}$,
or equivalently $t_{\sigma\alpha} = \frac{1}{2}(1-{\rm{Re}} S_{\sigma\alpha,\sigma\alpha})$, where the above convention $\nu=1/(2 \pi)$ was used.
At a FL FP, the scattering matrix is unitary,\cite{hewson}
$\boldsymbol{S}^{\dagger}\boldsymbol{S}=\boldsymbol{\mathds{1}}$, and
as such describes purely elastic scattering.
By contrast, at the NFL FP of the 2IK model, it has been
shown\cite{CFT2IKM} that $\boldsymbol{S}=\boldsymbol{0}$, which
implies fully inelastic scattering: a single electron sent in to
scatter off the impurities decays completely into collective
excitations, and no single-particle state emerges. Such behavior is
manifest by a half-unitarity spectrum,
$t_{\sigma\alpha} = \frac{1}{2}$.

However, along a crossover between FPs, the
Green function does not in general take the form of Eq.~\ref{cibc}.


\subsection{$S0(8)$ Majorana fermion representation}
\label{so8}

Further insight into the FPs of the 2IK model
is provided by a representation in terms of Majorana fermions
(MFs).
Considering again the free theory described by $H_0$, four nonlocal
fermions can be defined by Abelian bosonization and
refermionization\cite{emery,gan,maldacena} of the four original
Dirac fermions $\psi_{\sigma\alpha}$ with spin
$\sigma=\uparrow, \downarrow$ and channel index
$\alpha=L, R$. 8 MFs are then obtained by taking the real and imaginary
part of each.

Specifically, four bosonic fields $\phi_{\sigma\alpha}$
are defined, viz
 \begin{equation}
 \label{bosonization}
 \psi_{\sigma \alpha} \sim F_{\sigma \alpha} e^{-i \phi_{\sigma \alpha}},
  \end{equation}
where $F_{\sigma\alpha}$ are Klein factors.\cite{ZarandvonDelft} Linear
  combinations of these bosonic fields are then used to construct new
  fields,
  \begin{equation}
  \label{linear}
  \{ \phi_c,\phi_s,\phi_f,\phi_X\}=\tfrac{1}{2} \sum_{\sigma \alpha} \phi_{\sigma \alpha} \{1,(-1)^{\sigma+1},(-1)^{\alpha+1},(-1)^{\sigma+\alpha} \},
  \end{equation}
where $\sigma\equiv \uparrow,\downarrow=1,2$ and $\alpha\equiv
L,R=1,2$. Refermionizing yields,
\begin{equation}
\label{refermionization}
\psi_A \sim F_A e^{-i \phi_A},~~~ (A=c,s,f,X),
 \end{equation}
where the Klein factors $F_A$ are related to $F_{\sigma\alpha}$ as
described in Appendix~\ref{quadratic}. Thus, four new fermionic
species $\psi_A$ are defined, with $A=c,s,f,X$ corresponding to
`charge', `spin', `flavor' and `spin-flavor'. The real and imaginary
parts of each,
\begin{equation}
\label{MFdecomp}
\chi_1^A =\frac{\psi^\dagger_A + \psi_A}{\sqrt{2}},\qquad
\chi_2^A=\frac{\psi^\dagger_A - \psi_A}{\sqrt{2}i},
\end{equation}
fulfill the Majorana property $(\chi_i^A )^{\dagger}=\chi_i^A$ and
satisfy separately the fermionic anticommutation relation
$\{ (\chi_i^A)^{\dagger}(x),\chi_j^B (x')\} = \delta_{ij}\delta_{AB} \delta(x-x')$, and so are
referred to as MFs.

The free fermion FP Hamiltonian (corresponding to Eqs.~\ref{2ck} and
\ref{2ik} with $J=0$ and $\delta H=0$) then follows as
\begin{equation}
\label{MFfreeFP}
H_{FP}=\frac{i}{2}\int_{-\infty}^{\infty} \textit{d}x~
\vec{\chi}(x)\cdot \partial_x \vec{\chi}(x),
\end{equation}
where $\vec{\chi}\equiv \{
\chi_2^X,\chi_1^f,\chi_2^f,\chi_1^s,\chi_2^s,\chi_1^X,\chi_1^c,\chi_2^c
\}$, and the scattering states are defined by the trivial boundary
condition $\vec{\chi}(x)=\vec{\chi}(-x)|_{x\to 0}$. The fixed point thus
possesses a large $SO(8)$ symmetry in terms of these MFs.

Likewise, the FL FP Hamiltonian (in which the impurity degrees of
freedom are quenched) is similarly described by Eq.~\ref{MFfreeFP}.
The corresponding boundary condition is encoded in the single-particle
FL scattering S matrix $S_{\sigma\alpha,\sigma'\alpha'}$. Although it
depends on the specific perturbations generating the crossover, the
boundary condition is thus trivial at the FL FP. In particular, finite
detuning $K> K_c$ results in $\vec{\chi}(x)=\vec{\chi}(-x)|_{x\to 0}$, as
obtained at the free fermion FP.

The remarkable fact is that the NFL FP Hamiltonian also
takes the form of Eq.~\ref{MFfreeFP}; and its nontrivial boundary
condition\cite{CFT2IKM} is again simple
in terms of the MFs. It can be accounted for by defining the
scattering states $\chi_2^X(x)=-\chi_2^X(-x)|_{x\to 0}$, and
$\chi_j^A(x)=\chi_j^A(-x)|_{x\to 0}$ for $(j,A)\ne (2,X)$. Thus, 7 of the 8 MFs
are described by the free theory at the NFL FP.

\subsection{Bose-Ising decomposition}
\label{B-I}
The FP Hamiltonian Eq.~\ref{MFfreeFP} describes a higher $SO(8)$ symmetry
than is present in the original Hamiltonian. The explicit symmetries
of the 2IK model also allow a 
separation of the theory into different symmetry sectors. Specifically, the
$SU(2)_1 \times SU(2)_1 \times SU(2)_2\times Z_2$ symmetry sectors comprise a
Bose-Ising representation,\cite{CFT2IKM} describing a coset
construction of three Wess-Zumino-Witten (WZW) nonlinear $\sigma$
models, together with a $Z_2$ Ising model. The two $SU(2)_1$ theories
with central charge $c=1$ correspond to conserved charge in the left
and right channels.  The $SU(2)_2$ theory with $c=3/2$  
corresponds to conserved total spin.
Finally, the Ising
model $Z_2$ is a $c=1/2$ theory corresponding to a single MF.
This non-Abelian representation is connected with the 8 MFs, as
discussed in Ref.~\onlinecite{maldacena}. The symmetry `currents' of
those sectors, such as the spin current $\vec{J}(x)=\psi^\dagger(x)
\frac{\vec{\sigma}}{2} \psi(x)$, are represented \emph{quadratically} in terms
of MFs as described in Appendix~\ref{quadratic}. Specifically the
$SU(2)_1 \times SU(2)_1$ charge currents in left and right channels
are represented in terms of 4 MFs $\{\chi_1^f,\chi_2^f \}$ 
and $\{\chi_1^c,\chi_2^c \}$; while the $ SU(2)_2$ spin current
$\vec{J}$ is represented in terms of three MFs, 
$\vec{\chi}_s=\{\chi_1^s,\chi_2^s,\chi_1^X  \}$. The $Z_2$ theory corresponds to the remaining MF, $\chi_2^X$.

The important implication for our purposes is that the Green function
can be factorized into pieces coming from different sectors associated
with the various MFs. We now exploit the above
Bose-Ising construction,\cite{CFT2IKM} in terms of which the fermion
field can be expressed by the bosonization formula, 
\begin{equation}
\label{bosonizationf} \psi_{\sigma \alpha}(x) \propto [h_\alpha]_1(x) g_\sigma(x) \sigma_L(x).
 \end{equation}
 Here, the dimension $d=1/2$ fermion field has been decomposed into a
 dimension $d_h=1/4$ factor $[h_\alpha]_1$ representing the
 $\alpha=L,R$ spin-$\tfrac{1}{2}$ primary field of the $SU(2)_1$
 charge theories, a dimension $d_g=3/16$ factor $(g_\sigma)$
 representing the spin-$\tfrac{1}{2}$ primary field of the $SU(2)_2$
 spin theory, and the dimension $d_\sigma=1/16$ factor $\sigma_L$
 originating from the Ising sector. The subscript $L$ emphasizes
 that $\sigma_L$ is only the left-moving chiral component of the full
 spin operator $\sigma$ of the Ising sector, arising here because
 $\psi_{\sigma\alpha}$ is the chiral left-moving fermion 
 field. Since the NFL FP of the 2IK model
 is conformally invariant,\cite{CFT2IKM} we may use
 Eqs.~\ref{bosonizationf} and \ref{corrfuncT} to determine the
 contribution to the full Green function coming from each of the
 sectors:  
\begin{equation}
\label{factorG}
\begin{split}
G^{NFL}_{\sigma\alpha,\sigma'\alpha'}(\tau,x_1-x_2) \propto ~&\delta_{\sigma \sigma'} \delta_{\alpha \alpha'} \left [ G^{0}(\tau,x_1-x_2)
  \right ]^{\tfrac{7}{8}}\\
 &\times \langle \sigma_L(\tau,ix_1)\sigma_L(0,ix_2) \rangle.
\end{split}
\end{equation}
The free boundary condition in the charge and spin sectors yields the
first factor, corresponding to the free Green function Eq.~\ref{freeG}
but with power $2 (d_h+d_g)=\tfrac{7}{8}$ arising because 7 of the 8
MFs are associated with these sectors. The NFL boundary condition is
expressed in terms of fusing with the $d_{\sigma}=1/16$ field $\sigma_L$
from the Ising sector in Eq.~\ref{bosonizationf}. The second factor
thus comes from the remaining Ising sector. At the NFL FP it follows
from Eq.~\ref{corrfuncT} that   
$\langle \sigma_L(\tau,ix_1)\sigma_L(0,ix_2) \rangle = \left (
  \frac{S_{1/16}^{1/16} /S_0^{1/16}}{S_{1/16}^0 /S_0^0}\right )  \left
  [ G^{0}(\tau,x_1-x_2) \right ]^{\tfrac{1}{8}}$. Since the modular S
matrix for fusion with the $a=1/16$ Ising operator has a vanishing
element $S_{1/16}^{1/16}=0$, the entire Green function thus vanishes
at the NFL FP;\cite{CFT2IKM} consistent with Eq.~\ref{cibc} with
$S_{\sigma \alpha,\sigma' \alpha'}=0$.

In summary, the nontrivial boundary condition at the NFL FP
affects only the \emph{Ising} sector of the 2IK model. The
function  $\langle \sigma_L(\tau,ix_1)\sigma_L(0,ix_2) \rangle$ in
Eq.~\ref{factorG} is a quantity pertaining to the $Z_2$ Ising
model and contains the nontrivial physics; while the spin and charge
sectors are simply spectators. 
In the next section we exploit Eq.~\ref{factorG} and a connection
between the 2IK model and a classical Ising model\cite{CFT2IKM} to
determine the Green function along the \emph{crossover} from the NFL
FP to the FL FP.


\subsection{Crossover due to $K\ne K_c$ in the 2IK model}
\label{gft0}
As highlighted above, the NFL FP of the 2IK model is described by the
free theory in all sectors except the Ising sector, which takes a
modified boundary condition.
When the small perturbation $\lambda_1$ is included
(corresponding to detuning $K\ne K_c$), the NFL FP is
destabilized. The effective Hamiltonian in the vicinity of the NFL FP
is then $H_{QCP} = H_{FP}[\vec{\chi}]+\delta H_{QCP}$, with $H_{FP}$
the FP Hamiltonian itself, given in
Eq.~\ref{MFfreeFP} and parametrized in terms of the scattering states
$\vec{\chi}$ which encode the boundary condition. The correction is
given by,
\begin{equation}
\label{deltaHlambda1}
\delta H_{QCP} \propto i \lambda_1 \chi_2^X(0)a,
\end{equation}
where $a$ is a local MF involving impurity spin operators\cite{gan}
which satisfies $a^2=1$ and anticommutes with all other MFs,
$\chi_j^A$.

The perturbation $\lambda_1$ thus acts only in the Ising sector.
Indeed, the difference in boundary conditions between the NFL and FL
FPs is also confined to the Ising sector (all other sectors have free
boundary conditions at both FPs).  The \emph{entire crossover} from
NFL to FL FP thus occurs completely within the Ising sector because 
the $\lambda_1$ perturbation does not spoil the decoupling of
the sectors, which becomes exact\cite{SelaPRL} in the limit $T^*\ll
T_K$. The other 
sectors then act as spectators along the crossover.  

This implies a generalization of Eq.~\ref{factorG} to the
full crossover. 
One can still interpret the first factor in Eq.~\ref{factorG} with
power $7/8$ as the product of autocorrelation functions of 7 spin
fields which undergo no change in the boundary condition.  But the
$\sigma$ field originating from the $Z_2$ sector in the $SU(2)_1
\times SU(2)_1 \times SU(2)_2\times Z_2$ construction has
\emph{flowing} boundary condition (which is not conformally
invariant). 

We consider first the equal-time Green function with
$x_1=-x_2$ positioned symmetrically on either side of the boundary:
\begin{equation}
\label{factorGcrossover}
\begin{split}
\langle
&\psi_{\sigma\alpha}^{\phantom{\dagger}}(\tau,x)\psi_{\sigma'\alpha'}^{\dagger}(\tau,-x)
\rangle \equiv G_{\sigma\alpha,\sigma'\alpha'}(0,x,-x),\\
&\propto \left [ G^{0}(0,2x)
  \right ]^{\tfrac{7}{8}}  \times  \langle
\sigma_L(\tau,ix)\sigma_L(\tau,-ix) \rangle,
\end{split}
\end{equation}
where the factor
$\langle\sigma_L(\tau,ix)\sigma_L(\tau,-ix) \rangle$ now describes the
\emph{crossover} in the Ising sector in terms of the chiral
(left-moving) Ising magnetization operator $\sigma_L(\tau,ix)$.

\begin{figure}[t] \begin{center} \includegraphics*[width=85mm]{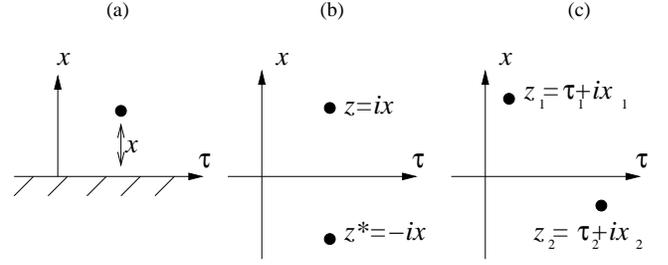}
    \caption{A one-point function of a bulk operator evaluated at
      distance $x$ from the boundary in (a) is mapped to the two point
      function of the associated chiral fields in the absence of a boundary shown in (b) and evaluated at image positions with respect to the line $x=0$. (c) Generalization of the two point function away from image points.
\label{fig:images}}
\end{center}
\end{figure}

We now use standard boundary CFT methods\cite{cardy} to relate the
two-point chiral function $\langle \sigma_L(\tau,ix)\sigma_L(\tau,-ix)
\rangle$ living in the full plane [see Fig. 9
(b)] to the product of chiral holomorphic $\sigma_L(\tau+ix)$ and
antiholomorphic $\sigma_R(\tau-ix)$ operators living in the halfplane
with a boundary.
But bulk operators in CFT may be
expressed as $\sigma(\tau,x)=\sigma_L(\tau+ix)\sigma_R(\tau-ix)$, and
so the desired correlator is simply
$\langle \sigma_L(\tau,ix)\sigma_L(\tau,-ix)
\rangle=\langle\sigma(\tau,x)\rangle$ in terms of the bulk Ising
magnetization operator, evaluated at a distance $x$ from the boundary
[see Fig.~\ref{fig:images} (a)]. Finally, we note that
$\sigma(\tau,x)\equiv \sigma(x)$ is independent of $\tau$ due
to translational invariance along the boundary. The Green function
along the NFL to FL crossover then follows as,
\begin{equation}
\label{factorGising}
\begin{split}
G_{\sigma\alpha,\sigma'\alpha'}(0,x,-x)=&\delta_{\sigma \sigma'} \delta_{\alpha \alpha'}\left ( \frac{1}{8\pi i} \right )^{\tfrac{1}{8}} \left [ G^{0}(0,2x)
  \right ]^{\tfrac{7}{8}}  \langle \sigma(x) \rangle,\\
\equiv& G_{\sigma\alpha,\sigma'\alpha'}(x),
\end{split}
\end{equation}
(with the factor $(8\pi i)^{-1/8}$ required for normalization). At
long distances where the Green function describes the FL FP,
Eq.~\ref{factorGising} implies $\langle \sigma(x) \rangle=(2/x)^{1/8}$ as
$T\rightarrow 0$.

We now exploit a connection\cite{CFT2IKM} between the 2IK model at
criticality with a simpler Ising model to obtain $\langle \sigma(x)
\rangle$ and hence the exact Green function
$G_{\sigma\alpha,\sigma'\alpha'}(x)$ along the NFL to FL crossover.

When a small magnetic field $h$ is applied to the boundary of a
classical Ising model on a semi-infinite plane, the local magnetization
shows a crossover\cite{cardy,Cardy91} as a function of distance from
the boundary. The crossover in this boundary Ising model (BIM) can be
understood as an RG flow\cite{cardy,Cardy91} from an unstable FP at short
distances (with free boundary condition $h=0$), to a stable FP at
large distances (with fixed boundary condition $h=\pm \infty$). The
universal crossover is characterized by an energy scale
$T^*\equiv 4\pi h^2$ (or a corresponding lengthscale $\xi^*\propto
1/T^*$).

Importantly, it was shown in Ref.~\onlinecite{CFT2IKM} that the RG
flow in the BIM due to small $h$ at the critical temperature is
identical to the RG flow in the 2IK model due to small detuning $K\ne
K_c$ at $T=0$. As such, the NFL and FL FPs of the 2IK model can be
understood in terms of the BIM FPs with free and fixed boundary
conditions. The crossover energy scale in the 2IK model can then be
identified as
\begin{equation}
\begin{split}
\label{bim2ikT*}
T^*&=\frac{c_1^2(K-K_c)^2}{T_K}\equiv \lambda_1^2\qquad\text{:~2IK}\\
&=4\pi h^2\qquad\qquad\qquad\qquad\text{:~BIM}
\end{split}
\end{equation}
with $c_1=\mathcal{O}(1)$ as in Table~\ref{table:pert}.

In Ref.~\onlinecite{cz} Chatterjee and Zamolodchikov derived an exact
expression for the Ising magnetization $\langle \sigma(x)
\rangle$ on the semi-infinite plane geometry in the continuum
limit. Their result\cite{cz} is
\begin{equation}
\label{czmag}
\langle \sigma(x) \rangle_{CZ} = \pm(2/ x)^{1/8} \sqrt{8  h^2 x}  ~e^{4 \pi h^2 x} K_0(4 \pi h^2 x),
\end{equation}
with $K_0$ the modified Bessel function of the second kind; and $\pm$
for $h \gtrless 0$. We take now $h>0$ (corresponding to $K>K_c$) for
concreteness. Note that Eq.~\ref{czmag} yields asymptotically $\langle \sigma(x) \rangle =
(2/x)^{1/8}$ at long distances $x\rightarrow \infty$, consistent with
the normalization of Eq.~\ref{factorGising}.

We now show that the analyticity of the Green function and the local
nature of the Kondo interaction implies a generalization of
$G_{\sigma\alpha,\sigma'\alpha'}(x)$ in terms of spatial coordinate
$x$ [Fig.~\ref{fig:images} (b)], to
$G_{\sigma\alpha,\sigma'\alpha'}(z_1-z_2)\equiv \langle
\psi_{\sigma\alpha}^{\phantom{\dagger}}(z_1)\psi_{\sigma'\alpha'}^{\dagger}(z_2)
\rangle$  in terms of general complex coordinates $z_1=\tau+ix_1$ and
$z_2=ix_2$ [Fig.~\ref{fig:images} (c)]. Using the free chiral Green
function Eq.~\ref{G0} in the definition of the t matrix Eq.~\ref{Tdefinition}, one
obtains
\begin{equation}
\label{tmatGmat}
\begin{split}
\mathcal{G}_{\sigma\alpha,\sigma'\alpha'}(x_1,x_2,i
\omega_n)=-&\delta_{\sigma\sigma'}\delta_{\alpha\alpha'}\theta(-\omega_n)\times\\
&e^{\omega_n (x_1-x_2)}[i + \mathcal{T}_{\sigma\alpha,\sigma'\alpha'}(i \omega_n)],
\end{split}
\end{equation}
where $x_1>0$ and $x_2<0$ as before, and the t matrix is local in
space.

The Matsubara transform then yields
\begin{equation}
\label{tmatz1z2}
\begin{split}
G_{\sigma\alpha,\sigma'\alpha'}(\tau,x_1,x_2)=-&
\frac{\delta_{\sigma\sigma'}\delta_{\alpha\alpha'}}{\beta}\sum_n
\theta(-\omega_n) \times\\
& e^{-i   \omega_n (\tau
  +ix_1-ix_2)}[i+\mathcal{T}_{\sigma\alpha,\sigma'\alpha'}
(i \omega_n) ]\\
\equiv & G_{\sigma\alpha,\sigma'\alpha'}\left (\frac{z_1-z_2}{2i}\right ).
\end{split}
\end{equation}
Thus, the Green function only depends on $(z_1-z_2)$. This is a
somewhat counter-intuitive result, because the boundary breaks the
translational invariance along the spatial coordinate $x$. In
Appendix~\ref{app_analyticity} we give an alternative proof, showing
that Eq.~\ref{tmatz1z2} holds to all orders in perturbation theory
around the NFL FP.

Comparing Eq.~\ref{tmatz1z2} and Eq.~\ref{factorGising}, it follows that
$G_{\sigma\alpha,\sigma'\alpha'}(0,x,-x)=G_{\sigma\alpha,\sigma'\alpha'}(\tau,x_1,x_2)$
when $x=(z_1-z_2)/(2i)$.  Employing this substitution in
Eqs.~\ref{factorGising} and \ref{czmag}, then taking the Matsubara
transform, we obtain
\begin{equation}
\label{analcont1}
\begin{split}
\mathcal{G}_{\sigma\alpha,\sigma'\alpha'}(x_1,x_2, i \omega_n) = &\delta_{\sigma\sigma'}\delta_{\alpha\alpha'}\int_{-\infty}^{\infty} d \tau ~
\left(\frac{e^{i \omega_n \tau}}{4 \pi i x}\right ) \times\\
&\sqrt{8 h^2 x}~e^{4 \pi h^2 x} ~K_0(4 \pi h^2 x) ,
\end{split}
\end{equation}
where we have used $\beta\rightarrow \infty$ as appropriate for
Eq.~\ref{czmag}, and so
$\omega_n=\tfrac{\pi}{\beta}(1+2n)$ is continuous. Setting $x_1=0^+$,
$x_2=0^-$, we define the infinitesimal $\delta= x_1 -
x_2>0$, such that $x=\frac{\tau+i\delta}{2i}$.
With the integral representation of the Bessel function $K_0(z) =
e^{-z} \int_0^\infty dk \frac{e^{-2 k z}}{\sqrt{k(k+1)}}$ (for
${\rm{Re}}~z>0$), we obtain
\begin{equation}
\label{intrep}
\begin{split}
\mathcal{G}_{\sigma\alpha,\sigma'\alpha'}(0^+,0^-, i \omega_n) =\frac{\delta_{\sigma\sigma'}\delta_{\alpha\alpha'}}{\pi \sqrt{ i}} h\int_0^\infty   dk~\frac{g(\omega_n+4 \pi h^2 k)}{ \sqrt{k(k+1)}},
\end{split}
\end{equation}
with the integral over $\tau$ evaluated by contour methods,
\begin{equation}
\label{tauint}
g(z)=\int_{-\infty + i \delta}^{\infty+i\delta} d \tau~\frac{e^{i
    z\tau}}{\sqrt{\tau}}  =\theta(-z)\sqrt{\frac{4\pi i}{z}}.
\end{equation}
For negative Matsubara frequencies $\omega_n<0$, the Green function then follows as
\begin{equation}
\label{T0matsG}
\begin{split}
\mathcal{G}_{\sigma\alpha,\sigma'\alpha'}&(0^+,0^-, i \omega_n) \\ &=\frac{\delta_{\sigma\sigma'}\delta_{\alpha\alpha'}}{\pi i} \int_0^{-\frac{\omega_n}{4 \pi h^2}} \frac{dk}{ \sqrt{k(k+1)(-\frac{\omega_n}{4 \pi h^2}-k)}}\\
& \equiv\delta_{\sigma\sigma'}\delta_{\alpha\alpha'}\frac{2 }{\pi i} K\left( \frac{\omega_n}{4 \pi h^2}\right),
\end{split}
\end{equation}
where the $k$ integral has been expressed more simply in the last line
in terms of the complete elliptic integral of the first kind, $K(z) = \int_0^{\pi/2} \frac{d \theta}{\sqrt{1-z \sin^2 \theta}} = \int_0^1 \frac{dt}{\sqrt{(1-t^2)(1-z t^2)}}$.

Since $K(z)$ has a branch cut discontinuity in the complex $z$-plane
running from $1$ to $\infty$, the analytic continuation to real
frequencies, $i \omega_n \to \omega + i0^+$ can be performed without
crossing any singularity.
Thus, if one has $G_{\sigma\alpha,\sigma'\alpha'}(x)$ as an \emph{analytic}
function of spatial coordinate $x$ (as in Eq.~\ref{factorGising}), then
a full knowledge of both space and time dependences of the Green
function is implied by analytic continuation.

Using Eq.~\ref{Tdefinition}, one recovers our earlier
result\cite{SelaMitchellFritz} for the $T=0$ crossover t matrix in the
2IK model due to perturbation $K\ne K_c$,
\begin{equation}
\label{TmatrixT=0}
2\pi i \nu \mathcal{T}_{\sigma\alpha,\sigma'\alpha'}(\omega,T=0) =
\delta_{\sigma\sigma'}\delta_{\alpha\alpha'}\left (1\mp\frac{2 }{\pi } K[- i  \omega/ T^*]\right) ,
\end{equation}
in terms of $T^*\equiv 4\pi h^2$ (and with $\nu=1/(2\pi)$ as
before). Here $\mp$ is used for $K\gtrless K_c$, corresponding to
local singlet or Kondo screened phases of the 2IK model,
respectively. Eq.~\ref{TmatrixT=0} is thus equivalent to
Eqs.~\ref{exactresult} and \ref{exactT=0} with the scattering S matrix
$S_{\sigma\alpha,\sigma'\alpha'}=\pm \delta_{\sigma\sigma'}\delta_{\alpha\alpha'}$.

In the next section, we generalize these results to finite temperature.


\section{Derivation at finite temperatures}
\label{finiteTderiv}

Our starting point for the derivation of the finite-temperature
crossover Green function is Eq.~\ref{factorGising}.
$G_{\sigma\alpha,\sigma'\alpha'}(x)$ thus follows from the
Ising magnetization $\langle \sigma(x,\beta) \rangle$ evaluated at
temperature $T\equiv \beta^{-1}$.

In Ref.~\onlinecite{ising_note} we considered the magnetization at
temperature $\beta^{-1}$ and distance $x$ from the boundary in a quantum 1D
transverse field Ising critical chain, with a magnetic field $h$ applied
to the first spin at the point boundary. It is given by
\begin{equation}
\label{finitetisingmag}
\langle \sigma(x,\beta) \rangle = f(\beta h^2)\times \langle \sigma(x,\beta) \rangle_{\textit{LLS}},
\end{equation}
with 
\begin{equation}
\label{ffunc}
f(\beta h^2)=\sqrt{2\beta h^2} ~\frac{\Gamma[\tfrac{1}{2} + 2\beta
  h^2]}{\Gamma[1+2\beta h^2]},
\end{equation}
and where
\begin{equation}
\label{lls}
\begin{split}
\langle \sigma(x,\beta) \rangle_{\textit{LLS}} =  &\left( \frac{\frac{4 \pi}{\beta}}{\sinh  \frac{2 \pi x}{\beta}}  \right)^{\tfrac{1}{8}} \times  \\
&_{2}F_1\left(\frac{1}{2},\frac{1}{2};1+2 \beta h^2,\frac{1-\coth \frac{2 \pi x}{\beta}}{2} \right),
\end{split}
\end{equation}
is the result of Leclair, Lesage and Saleur
in Ref.~\onlinecite{Leclair}, who generalized the $T=0$ result of
Chatterjee and Zamolodchikov\cite{cz} for the semi-infinite plane
geometry (Eq.~\ref{czmag})
to the geometry of a semi-infinite \emph{cylinder} with circumference
$\beta$, and magnetic field $h>0$ applied to the circular
boundary. The latter is equivalent to the quantum Ising chain with
transverse field. We showed\cite{ising_note} that while $\langle
\sigma(x,\beta) \rangle_{LLS}$ gives the full and highly nontrivial $x$ dependence
of $\langle \sigma(x,\beta) \rangle$, it misses the multiplicative scaling
function of the variable $\beta h^2$, given in Eq.~\ref{ffunc}. 

In the low-temperature limit $\beta\rightarrow \infty$, one obtains $f(\beta h^2)\rightarrow
1$ and
\begin{equation}
\label{limit}
\begin{split}
~_{2}F_1\left(\frac{1}{2},\frac{1}{2};1+2 \beta h^2,\frac{1-\coth \frac{2 \pi x}{\beta}}{2} \right) \overset{\beta \to \infty}{\to}  \\
 (8  h^2 x)^{1/2}   e^{4 \pi h^2 x} K_0(4 \pi h^2 x)   ,
\end{split}
\end{equation}
such that Eqs.~\ref{finitetisingmag}, \ref{lls} reduce as they must to
Eq.~\ref{czmag}. In the limit $h\rightarrow \infty$,  one recovers the
fixed boundary condition, describing the FL fixed point. Again $f(\beta h^2)\rightarrow
1$ but Eq.~\ref{lls} reduces now to
\begin{equation}
\label{largeh}
\langle \sigma(x,\beta) \rangle_{\textit{LLS}} \overset{h \to \infty}{\to} \left( \frac{\frac{4 \pi}{\beta}}{\sinh  \frac{2 \pi x}{\beta}}  \right)^{\tfrac{1}{8}}.
\end{equation}
This limiting behavior can also be obtained using
the conformal mapping from the semi-infinite plane geometry with boundary
${\rm{Im}}z =0$, to the semi-infinite cylinder geometry parametrized by
${\rm{Re}}z' \in (-\beta/2,\beta/2)$, with boundary
${\rm{Im}}z' =0$,
 \begin{equation}
 \label{confmap}
 z=\tan \left(\frac{\pi z'}{\beta} \right).
 \end{equation}
On the semi-infinite plane, the Ising magnetization in the limit
$h\rightarrow \infty$ is known\cite{Cardy91} to decay as $\langle \sigma(x) \rangle=(2/x)^{1/8}$,
yielding precisely Eq.~\ref{largeh} on the semi-infinite cylinder.

Combining Eqs.~\ref{factorGising}, \ref{finitetisingmag}--\ref{lls}
we obtain the crossover Green function at finite temperature,
\begin{equation}
\label{GfiniteTx}
\begin{split}
G_{\sigma\alpha,\sigma'\alpha'}(x)=\delta_{\sigma\sigma'}\delta_{\alpha\alpha'}\frac{\sqrt{2\beta h^2}}{2\beta
  i \sinh [\tfrac{2\pi x}{\beta}]}\times \\ \frac{\Gamma[\tfrac{1}{2} + 2\beta
  h^2]}{\Gamma[1+2\beta h^2]}~_{2}F_1\left(\frac{1}{2},\frac{1}{2};1+2 \beta h^2,\frac{1-\coth \frac{2 \pi x}{\beta}}{2} \right).
\end{split}
\end{equation}
For $h\rightarrow \infty$, this gives correctly
$G_{\sigma\alpha,\sigma'\alpha'}(0,x,-x)\equiv
G^{\textit{FL}}_{\sigma\alpha,\sigma'\alpha'}(0,x,-x)=1/(2\beta i \sinh [\tfrac{2  \pi
  x}{\beta}])$, as expected from the boundary CFT result
for the FL FP Green function,
Eq.~\ref{cibc} (with $S_{\sigma\alpha,\sigma\alpha}=1$).

Of course, the function $f(\beta h^2)$ does become important when
considering the behavior of the Green function over the entire range
of $\beta h^2$ (or equivalently, $T/T^*$). In particular, at  high
temperatures $\beta\rightarrow 0$ (and finite $h$), one obtains $f(\beta
h^2)\rightarrow 0$. Thus $G_{\sigma\alpha,\sigma'\alpha'}(0,x,-x)\equiv
G^{\textit{NFL}}_{\sigma\alpha,\sigma'\alpha'}(0,x,-x)=0$, again
correctly recovering the known boundary CFT result\cite{CFT2IKM} for
the NFL FP Green function, Eq.~\ref{cibc} (with
$S_{\sigma\alpha,\sigma'\alpha'}=0$). The factor $f(\beta h^2)$
is indeed necessary to cancel the unphysical logarithmic divergence of $\langle
\sigma(x,\beta) \rangle_{\textit{LLS}}$ as $\beta\rightarrow
0$ (see also Appendix~\ref{pertFL}).


\subsection{Ambiguities in analytic continuation at finite-$T$}
\label{ancont}
The quantity of interest is of course the t matrix
$\mathcal{T}(i\omega_n)$, related via
Eq.~\ref{Tdefinition} to the
Matsubara Green function, itself obtained from Eq.~\ref{GfiniteTx} via
\begin{equation}
\mathcal{G}_{\uparrow L,\uparrow L}(x_1,x_2,i \omega_n)
=\int_{-\beta/2}^{\beta/2} d\tau~
e^{i\omega_n\tau}~G_{\uparrow L,\uparrow L}\left ( \frac{z_1-z_2}{2i}\right ),
\end{equation}
with $z_1=\tau+ix_1$ and $z_2=ix_2$ as usual.  Using the
integral representation of the hypergeometric function,
\begin{equation}
\label{integralrep}
_{2}F_1(a,b;c;z)=\frac{\Gamma[c]}{\Gamma[b] \Gamma[c-b]} \int_0^1 dt ~\frac{t^{b-1}(1-t)^{c-b-1}}{(1-tz)^a},
\end{equation}
we then obtain
\begin{equation}
\mathcal{G}_{\uparrow L,\uparrow L}(0^+,0^-, i \omega_n) =\frac{\sqrt{2 \beta h^2}}{8 \pi^{3/2} i} \int_0^1 dt~ \frac{(1-t)^{2 h^2 \beta-\frac{1}{2}}}{\sqrt{t}} A_n,
\end{equation}
where
\begin{equation}
A_n = \int_{- \beta/2}^{\beta/2} d\tau~ e^{ \frac{i \pi
    \tau}{\beta}(2n+1)} \frac{\frac{\frac{4
      \pi}{\beta}}{\sinh \left(\frac{\pi}{\beta }(\delta - i \tau)
    \right)}}{\sqrt{1- t\frac{1- \coth \left(\frac{\pi}{\beta }(\delta - i \tau) \right)}{2}}},
\end{equation}
for $x_1=0^{+}$ and $x_2=0^{-}$ such that
$\delta=x_1-x_2>0$; and with $n= \tfrac{\beta\omega_n}{2\pi} -
\tfrac{1}{2}$ a negative integer. Using contour
integration, it can be shown that
\begin{equation}
\begin{split}
A_n= 8 \pi^{3/2} (-1)^{n+1} \frac{(1-t)^{-n-1}}{\Gamma(-n) \Gamma(\frac{3}{2}+n)}   \times  \\
 ~_{2}F_1\left(\frac{1}{2},1+n,\frac{3}{2}+n,\frac{1}{1-t}\right).
\end{split}
\end{equation}
However, the naive substitution $i \omega_n \to \omega$ (or $n \to \frac{\beta \omega}{2 \pi i}-
\frac{1}{2} $) is problematic here and leads
to unphysical divergences. Indeed, such analytic continuation always
involves ambiguities due to the fact that $(-1)^{2 n }=1$ on the
integers, but it becomes $- e ^{\beta \omega}$ upon analytic
continuation. Thus, it is hard to find the function
$\mathcal{G}_{\uparrow L,\uparrow L}(0^+,0^-, i \omega_n)$ which gives the physical analytic
continuation.


\subsection{Finite-$T$ Green function from\\Friedel oscillations}
\label{friedel}

Eq.~\ref{GfiniteTx} describes the chiral electron Green function
$G_{\sigma\alpha,\sigma'\alpha'}(x)\equiv \langle \psi_{\sigma
  \alpha}(x) \psi_{\sigma' \alpha'}^\dagger(-x) \rangle$. The
information contained in such Green functions is directly linked to
the physical density oscillations around impurity (Friedel
oscillations), which in turn can be calculated from the t matrix,
$\mathcal{T}(\omega)$.\cite{Mezei,affleck,akm:realspace}
Indeed, in Ref.~\onlinecite{akm:realspace} the real-space densities
and hence Green function $G_{\sigma\alpha,\sigma\alpha}(x)$ for the
NFL to FL crossover in the 2CK model was explicitly calculated at
$T=0$ using the exact t matrix announced in
Ref.~\onlinecite{SelaMitchellFritz}.  It was
also highlighted\cite{akm:realspace} that far from the impurity, the
integral transformation relating the t matrix to the Friedel
oscillations can be inverted.

In this section we exploit these connections to calculate
$\mathcal{T}(\omega)$ directly from the density
oscillations described by $G_{\sigma\alpha,\sigma\alpha}(x)$, and thus
circumvent the need for problematic analytic
continuation.


For simplicity we restrict ourselves to 1D in this section;
although we note that the resulting t matrix is general, because at
low energies the standard flat band situation of most interest is
recovered. The density of the 1D fermion field $\Psi(x)$ at position $x$ is
given by $\rho(x) = \langle \Psi^\dagger(x) \Psi(x) \rangle
$. Expanding around the left $(l)$ and right $(r)$ Fermi points at low
energies using Eq.~\ref{chiralft}, one obtains
\begin{equation}
\label{expansion}
\Psi(x) = \psi_r(x) e^{i k_F x} +  \psi_l(x) e^{-i k_F x} ,
 \end{equation}
with the oscillating part of the density following as
\begin{equation}
\label{rhoexp}
\rho_{{\rm{osc}}}(x) =\tfrac{1}{2} e^{2 i k_F x}  \langle \psi_l^\dagger(x) \psi_r(x) -\psi_r(x) \psi_l^\dagger(x) \rangle + {\rm{H.c.}}.
\end{equation}
In the presence of particle-hole symmetry, $\rho(x) = 1/2$ for all $x$ (with lattice spacing set to unity);
and there are no density oscillations, $\rho_{{\rm{osc}}}(x)
=0$. However, introduction of potential scattering breaks
particle-hole symmetry and leads generically to real-space density
oscillations, which contain information about the t matrix.
Such potential scattering produces a phase-shift $\delta_P$ at the
Fermi energy, independent of the underlying Kondo physics, but which
does modify the boundary condition at $x=0$, according to $\psi_r(0)
=e^{ 2 i \delta_P } \psi_l(0)$.
As before, we define a chiral left-moving field on the infinite line,
but now take into account this change in the boundary condition:
\begin{equation}
\label{chiralfps}
\psi(x) = \psi_l(x) \theta(x) + e^{-2 i \delta_P} \psi_r(-x)\theta(-x).
\end{equation}

Using the imaginary time ordering of the chiral Green function defined
in Eq.~\ref{factorGising}, we have $G(-x+i0^{+})=\langle
\psi(-x)\psi^{\dagger}(x)\rangle$ and $G(-x-i0^{+})=-\langle
\psi^{\dagger}(x)\psi(-x)\rangle$, from which it follows that
\begin{equation}
\label{denosc}
\begin{split}
\rho_{{\rm{osc}}}(x) =-\tfrac{1}{2} e^{2 i k_F x+2 i \delta_P}\Big [& G_{\sigma\alpha,\sigma\alpha}(-x- i
  0^{+})\\ &+G_{\sigma\alpha,\sigma\alpha}(-x + i 0^{+})\Big ]  + {\rm{H.c.}}.
\end{split}
\end{equation}

The oscillating part of the density given by Eq.~\ref{denosc} can also
be obtained\cite{Mezei,affleck,akm:realspace} from the t
matrix. Generalizing to finite temperatures, we have
\begin{equation}
\label{denosctps}
\begin{split}
\rho_{{\rm{osc}}}(x) = -\tfrac{1}{\pi} \int_{-\infty}^{\infty}
& d\omega~f(\omega,T)\times \\
&\text{Im}\left \{
  [G^0(\omega,x)]^2
  \mathcal{T}_{\sigma\alpha,\sigma\alpha}(\omega,\delta_P) \right \},
\end{split}
\end{equation}
where $f(\omega,T)$ is the Fermi function,
$G^0(\omega,x)$ is the free Green function
Eq.~\ref{freeG} as a function of real frequency $\omega$; and
$\mathcal{T}_{\sigma\alpha,\sigma\alpha}(\omega,\delta_P)$ is the
scattering t matrix, defined in the presence of the potential
scattering. As shown in Ref.~\onlinecite{affleck}, at low energies
\begin{equation}
\label{tmatps}
2\pi \nu \mathcal{T}_{\sigma\alpha,\sigma\alpha}(\omega,\delta_P)=e^{2
  i \delta_P} \Big [2\pi \nu
\mathcal{T}_{\sigma\alpha,\sigma\alpha}(\omega) + i \Big ] -i
\end{equation}
in terms of the desired t matrix defined \emph{without} the potential
scattering. Indeed, far from the impurity one obtains asymptotically\cite{affleck}
\begin{equation}
\label{g0largex}
[G^0(\omega,x)]^2=-\tfrac{1}{v_F^2}e^{2i
  k_F x + 2i \omega x /v_F}.
\end{equation}

The oscillating part of the density can then be expressed as
\begin{equation}
\label{denosct}
\begin{split}
\rho_{{\rm{osc}}}(x) = \frac{1}{4\pi v_F^2} \int_{-\infty}^{\infty}
& d\omega~[1-2f(\omega,T)] e^{2i k_Fx +2i \delta_P + 2i\omega x/v_F } 
\\
& \times\Big [i \mathcal{T}_{\sigma\alpha,\sigma\alpha}(\omega) -\tfrac{1}{2\pi\nu} \Big ] + \text{H.c.}.
\end{split}
\end{equation}
Comparing now to Eq.~\ref{denosc} and inverting the Fourier transform
by operating on the resulting equations with $\int_{-\infty}^\infty
\frac{d x}{ \pi} e^{- 2 i \omega x}$, we obtain
\begin{equation}
\label{resultT}
\begin{split}
2\pi i \nu \mathcal{T}_{\sigma\alpha,\sigma\alpha}(\omega) = &1 -
\frac{4\pi\nu v_F^2}{\tanh(\tfrac{\beta \omega}{2})} \int_{-\infty}^{\infty} dx~  e^{2 i \omega x/v_F} \times \\ &[G_{\sigma\alpha,\sigma\alpha}(x- i 0^{+})
+G_{\sigma\alpha,\sigma\alpha}(x + i 0^{+})],
\end{split}
\end{equation}
where $G_{\sigma\alpha,\sigma\alpha}(x)$ is given as an analytic
function in Eq.~\ref{GfiniteTx}. We now set $v_F\equiv 1$ and
$2\pi\nu\equiv 1$ as before. Note that the hypergeometric
function $_{2}F_1(a,b,c,z)$ has a branch cut discontinuity in the
complex $z$ plane running from $1$ to $\infty$. The discontinuity
occurs only in the imaginary part of the function, with
${\rm{Im}}~_{2}F_1(a,b,c,z + i 0^{+}) =- {\rm{Im}}~_{2}F_1(a,b,c,z - i 0^{+})
$ for $z > 1  $. Furthermore ${\rm{Im}}~_{2}F_1(a,b,c,z )=0$ for $z \le
1$. Thus, integrating symmetrically above and below the real $x$ axis,
as per Eq.~\ref{resultT}, amounts to taking only the \emph{real} part of
$_{2}F_1(a,b,c,z)$; whence we obtain our final result
\begin{equation}
\label{result}
\begin{split}
&2\pi i \nu \mathcal{T}_{\sigma\alpha,\sigma\alpha}(\omega) = 1
+\frac{2 i \sqrt{2\beta h^2}}{\tanh\left ( \tfrac{\beta\omega}{2}\right )}\frac{\Gamma(\tfrac{1}{2}+2\beta h^2)}{\Gamma(1+2\beta
  h^2)} \int_{-\infty}^{\infty} dx~  \times \\
&
\left(  \frac{e^{2 i \omega x}
  }{\beta \sinh \frac{2 \pi  x }{\beta}} \right){\rm{Re}}~_{2}F_1\left(\frac{1}{2},\frac{1}{2};1+2 \beta h^2,\frac{1-\coth \frac{2 \pi x}{\beta}}{2} \right).
\end{split}
\end{equation}
Using the definition of the crossover scale $T^* = 4 \pi h^2$ (see
Eq.~\ref{bim2ikT*}), this gives the announced result
Eqs.~\ref{exactresult}, \ref{analyticformula} for the 2IK model in the
special case of perturbation $K>K_c$, where
$S_{\sigma\alpha,\sigma'\alpha'}=\delta_{\sigma\sigma'}\delta_{\alpha\alpha'}$. By
simple extension, for
$K<K_c$ (corresponding to $h<0$) one obtains the same crossover but
with
$S_{\sigma\alpha,\sigma'\alpha'}=-\delta_{\sigma\sigma'}\delta_{\alpha\alpha'}$.
Finally, we
note that taking the limit $\beta\rightarrow \infty$ of
Eq.~\ref{result} yields correctly Eq.~\ref{TmatrixT=0}.

In the next section we show that an \emph{identical} crossover occurs
in the 2CK model due to channel anisotropy.


\section{Crossover in the 2CK model}
\label{2ckgen}

The NFL fixed point Hamiltonians of the 2IK model and the 2CK model
have the same basic
structure.\cite{gogolin,emery,maldacena,Zarand2006,akm_2ckin2ik}
Although the underlying symmetries of the 2CK model are different from
those of the 2IK model, the free conduction electron Hamiltonian can
be written in terms of the same MFs in both cases (see Eqs.~\ref{bosonization}--\ref{MFdecomp}).
The CFT decomposition\cite{CFT2CK} of the 2CK model into  $U(1)
\times SU(2)_2 \times SU(2)_2$ symmetry sectors (corresponding to
conserved charge, spin and flavor), can then be expressed in terms of
these MFs: the $U(1)$ theory with central charge $c=1$ consists of a
free boson or equivalently two MFs $\chi_j^c~~(j=1,2)$; the spin
$SU(2)_2$ theory with $c=\frac{3}{2}$ consists of three MFs $\vec{\chi}_s =
(\chi_1^s,\chi_2^s,\chi_1^X)$; similarly the flavor $SU(2)_2$ theory
consists of three MFs $\vec{\chi}_f =
(\chi_2^f,-\chi_1^f,-\chi_2^X)$. The charge, spin and flavor currents
can also be written in terms of the MFs corresponding to those symmetry
sectors, as given in Eq.~\ref{Jchi2} of
Appendix ~\ref{quadratic}.

In particular, the NFL fixed point Hamiltonian is of the form of
Eq.~\ref{MFfreeFP}, with a boundary condition that is again
simple in terms of the MFs. In the 2CK model, the
NFL physics arises due to a modification of the boundary condition in
the spin sector only (the free boundary condition pertains in charge and
flavor sectors). The nontrivial boundary condition can be accounted
for by defining the scattering states $\vec{\chi}_s(x)=
-\vec{\chi}_s(-x)$ and $\vec{\chi}_f(x)=
\vec{\chi}_f(-x)$, $\chi_j^c(x)=\chi_j^c(-x)$ for $j=1,2$.
Indeed, the finite-size spectrum at the NFL FP\cite{CFT2CK} can be
understood in terms of excitations of a free Majorana chain.\cite{hewsonO3}

The NFL fixed point of the 2CK model is destabilized by certain
symmetry-breaking perturbations. These
perturbations can again be matched to MFs, with
the correction to the NFL fixed point Hamiltonian being of the form of
Eq.~\ref{deltaHlambda1} in the simplest case of channel anisotropy
$\lambda_1\propto \Delta_z\ne 0$ (see Table~\ref{table:pert}).\cite{CFT2CK}

Importantly, it was shown recently in Ref.~\onlinecite{akm_2ckin2ik} that
the NFL fixed
points of the 2CK and 2IK model are in fact \emph{identical} in the
sense that they both lie on the same marginal manifold parametrized by
potential scattering. Indeed, the low-energy effective model for the
2IK model in the limit of strong channel asymmetry is the 2CK
model,\cite{Zarand2006,akm_2ckin2ik}
but with an additional $\pi/2$ phase shift felt by the conduction
electrons of one channel.\cite{akm_2ckin2ik} For
concreteness, we consider now a variant of the standard 2IK
model in which channel asymmetry appears explicitly:
\begin{equation}
\label{asym2ik}
H_{2IK}=H_0+J_L\vec{S}_L.\vec{s}_{0L} + J_R\vec{S}_R.\vec{s}_{0R} + K \vec{S}_L.\vec{S}_R.
\end{equation}
One thus recovers Eq.~\ref{2ik} at the symmetric point $J_L=J_R$. In
the limit $J_L\gg
J_R$, the left impurity is Kondo screened by the left lead on the
single-channel scale $T_K^L$. At energies $\sim T_K^L$ the right
impurity is still essentially free. However, it feels a renormalized
coupling to its attached right lead, and an effective coupling to the
remaining Fermi liquid baths states of the left lead (which suffer a
full $\pi/2$ phase shift due to the first-stage Kondo effect in that
channel). The relative effective coupling strengths between left and
right channels can be tuned by the interimpurity coupling $K$. Tuning
$K$ to its critical value $K_c$ yields the 2CK critical
point;\cite{Zarand2006}  while
deviations $K\ne K_c$ correspond to finite channel anisotropy
$\Delta_z\ne 0$ in the effective 2CK model. In consequence, one
expects the NFL to FL crossover in the two models to be simply
related.

Since the NFL fixed point itself is the same in both 2CK and 2IK
models (up to potential scattering),\cite{akm_2ckin2ik} and because
the correction to the fixed point Hamiltonian due to the $\lambda_1$
perturbation is the same,\cite{CFT2CK,CFT2IKM} the RG flow along the
NFL to FL crossover is identical. To calculate the corresponding
crossover Green function, we simply incorporate the additional $\pi/2$
phase shift felt by the left-channel conduction electrons into our
scattering states definition. Using
$\psi_{\sigma L}' (x)= {\rm{sign}}(x)  \psi_{\sigma L}$, one
straightforwardly obtains
\begin{equation}
\label{2ik2ckcorr}
\begin{split}
\langle \psi_{\sigma \alpha}(z_1) \psi^\dagger_{\sigma' \alpha'}(z_2) \rangle_{2CK, \Delta_z>0}
& =\langle \psi'_{\sigma \alpha}(z_1) {\psi'}^\dagger_{\sigma' \alpha'}(z_2) \rangle_{2IK, K>K_c} \\
& =- \tau^z_{\alpha \alpha'} G_{\sigma \alpha,\sigma' \alpha'}\left( \frac{z_1 - z_2}{2i} \right),
\end{split}
\end{equation}
in terms of the analytic crossover Green function for the 2IK model
given in Eq.~\ref{GfiniteTx}. Following the steps of
Sec.~\ref{friedel}, the t matrix follows as
\begin{equation}
\label{2ik2cktmat}
2\pi i \nu \mathcal{T}_{\sigma\alpha,\sigma\alpha}^{2CK}(\omega) = 1-
 \tau^z_{\alpha \alpha'} [2\pi i \nu \mathcal{T}_{\sigma\alpha,\sigma\alpha}^{2IK}(\omega)-1].
\end{equation}

This result was obtained at $T=0$ in
Ref.~\onlinecite{SelaMitchellFritz}, where the
correspondence was checked by explicit numerical renormalization group
calculation.

In the 2CK model with $\Delta_z>0$, the left lead is more strongly
coupled, and completely screens the impurity at the FL FP on the
lowest energy scales; while the right lead decouples
asymptotically. The physical interpretation of
Eq.~\ref{2ik2cktmat} is thus that a Kondo resonance appears in the
spectral function on the scale of $T^*$ in the $\alpha=L$ channel,
while the resonance is destroyed in the $\alpha=R$ channel (hence the
dependence on the flavor-space Pauli matrix $\tau^z_{\alpha \alpha'}$).


\section{Generalization to arbitrary combination of perturbations}
\label{gengf}

In Sec.~\ref{finiteTderiv} we considered the finite-temperature crossover Green
function in the 2IK model due to the detuning perturbation $K\ne K_c$; while in
Sec.~\ref{2ckgen} we calculated the analogous crossover Green function in
the 2CK model due to channel anisotropy $\Delta_z\ne 0$. In this
section we generalize the results to an arbitrary combination of
perturbations in either model.

\subsection{Flavor rotation in the 2CK model}
\label{flavorrot}
Before discussing the full calculation, we motivate the general
approach by exploiting a bare symmetry of the 2CK model, in a simple
intuitive example.

Unlike the 2IK model, the 2CK model possesses a bare \emph{flavor}
symmetry (see Eq.~\ref{2ck}). The perturbations
$\Delta_x$, $\Delta_y$ and $\Delta_z$ break this symmetry, but are
themselves related by rotations in flavor-space.

A canonical transformation of the conduction electron operators of the bare
Hamiltonian is defined, viz
\begin{equation}
\label{cantrans}
\left ( \begin{array}{c} \psi_{k \sigma A} \\ \psi_{k \sigma B} \end{array} \right )
= \mathcal{U}
\left  ( \begin{array}{c} \psi_{k \sigma L} \\ \psi_{k \sigma R} \end{array} \right ),
\end{equation}
such that the unitary matrix  $\mathcal{U}$ satisfies 
$\mathcal{U} \left ( \vec{\Delta} \cdot \vec{\tau} \right ) \mathcal{U}^\dagger = |\Delta| \tau^z$. 
With the parametrization $\vec{\Delta} = |\Delta| (\sin \theta  \cos
\phi,\sin \theta  \sin \phi, \cos \theta)$, one obtains explicitly $\mathcal{U}=\exp \left( \frac{\theta}{2} (- \sin \phi \tau^x+\cos \phi \tau^y) \right)$. It follows that $\delta
H_{2CK}(\Delta_x,\Delta_y,\Delta_z)\rightarrow \delta
H_{2CK}(0,0,\tilde{\Delta}_z)$, with
$\tilde{\Delta}_z^2=\Delta_x^2+\Delta_y^2+\Delta_z^2$ (cf Eq.~\ref{tstar}). The physical
interpretation is that the impurity couples more strongly to one
linear combination of channels than the other. Thus the
perturbations $\lambda_1$, $\lambda_2$ and $\lambda_3$ in
Table~\ref{table:pert} are simply related, and their combined effect
enters only through $\vec{\lambda}_f$. In particular this implies only one fitting parameter $c_1$  for the
different components of the perturbation $\vec{\Delta}$ in the 2CK model.

The physical behavior in the case of arbitrary $\vec{\lambda}_f$ can
now be understood in terms of the situation where $\lambda_1$ alone
acts using the flavor rotation Eq.~\ref{cantrans}. For example, the
Green function $\langle\langle\psi_{k\sigma L};
\psi^\dagger_{k'\sigma L} \rangle\rangle_{\omega}$ probes the physical
channel $\sigma\alpha$ with $\alpha=L$ in the original basis. Using
the transformation Eq.~\ref{cantrans} it can be expressed as
\begin{equation}
\begin{split}
\langle\langle\psi_{k\sigma L};
\psi^\dagger_{k'\sigma L} \rangle\rangle_{\omega} =
 & \tfrac{1}{2}\left (1+\tfrac{\Delta_z}{\tilde{\Delta}_z} \right )
  \langle\langle\psi_{k\sigma A}; \psi^\dagger_{k'\sigma A} \rangle\rangle_{\omega}\\
+ &\tfrac{1}{2}\left (1-\tfrac{\Delta_z}{\tilde{\Delta}_z} \right )
  \langle\langle\psi_{k\sigma B}; \psi^\dagger_{k'\sigma B}
  \rangle\rangle_{\omega},
\end{split}
\end{equation}
in terms of the Green functions in the rotated basis, which correspond
to those calculated for $\tilde{\Delta}_z$ only, as considered in the
previous section. It is then easy to show that the t matrix for
arbitrary $\vec{\lambda}_f$ is given by
\begin{equation}
2\pi i \nu \mathcal{T}_{\sigma\alpha,\sigma\alpha}(\omega)= 1 + \left
  | \tfrac{\lambda_1}{\lambda} \right | (2 \pi i \nu
\tilde{\mathcal{T}}_{\sigma\alpha,\sigma\alpha}(\omega)-1)
\end{equation}
in terms of the t matrix $\tilde{\mathcal{T}}$ due to $\lambda_1$
given in Eq.~\ref{2ik2cktmat}. The simple rescaling of the spectral
function discussed in Sec.~\ref{results} and the precise form of
Eq.~\ref{s2ck} follow immediately.


\subsection{Emergent symmetries}

In this section we make use of the field theoretical description of the
NFL fixed point for both 2CK and 2IK models in the presence of
relevant perturbations.\cite{SelaPRL,Sela2009PRB,SelaPRL,Selemalecki}
A large $SO(8)$ \emph{emergent} symmetry at
the fixed point allows  these perturbations to be related by a unitary
transformation, in full analogy to the method demonstrated explicitly
in the previous section for the case of the bare flavor symmetry in
the 2CK model.

We express the NFL fixed point Hamiltonian in terms the
free MF scattering states,
\begin{equation}
\label{HQCP}
H_{QCP} = H_{FP}[\vec{\chi}] + \delta H_{QCP},
\end{equation}
where $H_{FP}[\vec{\chi}]$ is given in Eq.~\ref{MFfreeFP}, and with
$\vec{\chi}$ a vector of the 8 MFs. As already commented, the
structure of the NFL fixed point Hamiltonian is the same for 2CK and
2IK models; only the definition to the
scattering states is different. To fix notation, we define
\begin{equation}
\label{list}
\{ \chi_1,...\chi_8 \}=\{- {\rm{sign}(x)} \chi_2^X , \chi_1^f,\chi_2^f, \chi_1^s,\chi_2^s,\chi_1^X, \chi_1^c,\chi_2^c\}
\end{equation}
for the 2IK model; and
\begin{equation}
\label{list1}
\begin{split}
&\{ \chi_1,...\chi_8 \} =\\
& \{- \chi_2^X , \chi_2^f,-\chi_1^f, {\rm{sign}(x)} \chi_1^s,  {\rm{sign}(x)} \chi_2^s, {\rm{sign}(x)} \chi_1^X, \chi_1^c,\chi_2^c\}
\end{split}
\end{equation}
for the 2CK model. With this ordering of components, the correction to
the NFL fixed point due to relevant perturbations is given by
\begin{equation}
\label{deltaH}
\delta H_{QCP} \propto i \sum_{j=1}^8 \lambda_j \chi_j(0) a,
\end{equation}
with $a$ a local MF as before. The $\lambda_1$ perturbation
corresponding to $K\ne K_c$ in the 2IK model or $\Delta_z\ne 0$ in the
2CK model was considered explicitly in Eq.~\ref{deltaHlambda1}. The
other coupling constants are defined in Table~\ref{table:pert}; with
the single resulting crossover energy scale being the sum of their
squares, Eq.~\ref{tstar}. For a detailed derivation of
Eq.~\ref{deltaH}, see Refs.~\onlinecite{SelaPRL,Sela2009PRB,gan}. The
two MFs $\chi_7, \chi_8$ corresponding to the real and imaginary parts
of the total charge fermion, are not in fact allowed in the 2CK and
2IK models due to charge conservation.

\subsection{Unitary transformations}
\label{unitarytrans}

The crucial observation following from Eq.~\ref{deltaH} is that only
the linear combination $\lambda^{-1} \sum_{j=1}^8 \lambda_j \chi_j(x)$
of the 8 MF scattering states participates in the crossover. The
particular linear combination depends on the ratios of the various
perturbations (for example $K-K_c$, $V_{LR}$, $\vec{B}_s$ in the 2IK
model).

From Secs.~\ref{finiteTderiv} and \ref{2ckgen}, we know the crossover
Green function caused by the $\lambda_1$ perturbation. The strategy is
thus to fix the $\lambda_1$ perturbation as the direction in the
8-dimensional space of perturbations along which the Green function is
known, then use an $SO(8)$ rotation to obtain the general crossover
Green function.

We search now for a unitary operator $U U^\dagger = 1$ that transforms the
full Hamiltonian with an arbitrary combination of perturbations into
one involving the single perturbation $\lambda_1$. Specifically we
demand that,
\begin{equation}
\label{rot}
\begin{split}
U H_{FP}[\vec{\chi}]  U^\dagger &=H_{FP}[\vec{\chi}],\\
U \delta H_{QCP}  U^\dagger &=i \lambda \chi_1(0) a.
\end{split}
\end{equation}
This transformation is accomplished by an operator that rotates the
8-component vector $\vec{\chi}$ in the 8-dimensional space of
perturbations. The 28 generators of such rotations are of the form $i
\int dx \chi_j(x) A_{jj'} \chi_{j'}(x)$, where $A_{jj'}$ is a real
antisymmetric $8 \times 8$ matrix. It is easy to verify that the desired
operator satisfying Eq.~\ref{rot} is
\begin{equation}
\label{u}
U=e^{\theta \int_{-\infty}^\infty dx \chi_1(x) \chi_\perp(x)},
\end{equation}
where
\begin{equation}
\begin{split}
& \theta=\arcsin \frac{\lambda_\perp}{\lambda},\\
& \lambda_\perp=\sqrt{\lambda^2-\lambda_1^2},\\
& \chi_\perp(x) =\lambda_\perp^{-1} \sum_{j \ne 1} \lambda_j \chi_j(x).
\end{split}
\end{equation}

One can apply this transformation to the expectation value
of an operator written in terms of the original electrons, such as the
Green function $\langle \psi_{\sigma \alpha}(x) \psi^\dagger_{\sigma' \alpha'}(x')
\rangle_{H_{QCP}} = \langle U \psi_{\sigma \alpha}(x) U^{\dagger} U
\psi_{\sigma' \alpha'}^\dagger(x') U^\dagger \rangle_{U H_{QCP}
  U^\dagger}$. The crucial property of the unitary transformation
Eq.~\ref{u} is that it acts as a simple rotation also on the electron
fields. This occurs due to the existence of linear relations between
the 28 quadratic forms of the original electron fields and of the MFs
$\chi_{j}^A$, as discussed in Appendix~\ref{quadratic} and Ref.~\onlinecite{fisherbalents}.

As a simple relevant example, consider the 2IK model, perturbed by a
combination of $K-K_c$ and tunneling $V_{LR}$, such that only
$\lambda_1$ and $\lambda_2$ are finite. In this case the unitary
operator reads $U=e^{\theta \int_{-\infty}^\infty dx \chi_1(x)
  \chi_2(x)}$ with $\lambda_\perp = \lambda_2$ and $\lambda =\sqrt{
  \lambda_1^2 + \lambda_2^2}$. Using Eq.~\ref{list}, $\chi_1(x) \chi_2(x)=-{ \rm{sign}}(x) \chi_2^X(x)
\chi_1^f(x)$. The quadratic form $\chi_2^X(x)
\chi_1^f(x)$ is related to a quadratic form for the
original electrons $ \chi_2^X(x) \chi_1^f(x)=- \frac{i}{2}
\psi^\dagger \tau^x \psi$, as shown in Appendix~\ref{quadratic}.
The operator $U$ can now be understood as a simple rotation of
electron fields,
\begin{equation}
\label{M}
U \psi_{\sigma \alpha}(x) U^{\dagger} =\sum_{\sigma' \alpha'} M^{\gtrless}_{\sigma \alpha, \sigma' \alpha'}  \psi_{\sigma' \alpha'}(x),
\end{equation}
for $x \gtrless 0$; and where the rotation matrix acts here in flavor space,
\begin{equation}
\label{M1}
\left(M^{\gtrless}_{\sigma \alpha, \sigma' \alpha'}\right)_{\lambda_1,\lambda_2 \ne 0} =\delta_{\sigma \sigma'} \left[ \delta_{\alpha \alpha'}  \cos  \frac{\theta}{2}
\mp i   \tau^x_{\alpha \alpha'}  \sin  \frac{\theta}{2}  \right].
\end{equation}
The Green function then follows as
\begin{equation}
\begin{split}
& \langle \psi_{\sigma \alpha}(x) \psi^\dagger_{\sigma' \alpha'}(x')
\rangle_{H_{QCP}} = \sum_{\sigma_1 \alpha_1 \sigma'_1 \alpha'_1} \\
& M^>_{\sigma \alpha,\sigma_1 \alpha_1} ({M^<}^{\dagger})_{\sigma_1'
  \alpha_1',\sigma' \alpha'} \times \langle \psi_{\sigma_1
  \alpha_1}(x) \psi^\dagger_{\sigma_1' \alpha_1'}(x') \rangle_{U H_{QCP}
  U^{\dagger}},
\end{split}
\end{equation}
where $x>0$ and $x'<0$ is assumed. In terms of complex coordinates
$z_1 = \tau + i x_1$ and $z_2 = i x_2$ (with $x_1>0$, $x_2<0$),
the full Green function $\langle \psi_{\sigma \alpha}(z_1) \psi^\dagger_{\sigma' \alpha'}(z_2)
\rangle_{H_{QCP}}$ is then obtained from
$G_{\sigma_1\alpha_1,\sigma_1'\alpha_1'}(\frac{z_1-z_2}{2i})=-\langle
\psi_{\sigma_1 \alpha_1}(z_1) \psi^\dagger_{\sigma_1' \alpha_1'}(z_2)
\rangle_{U H_{QCP} U^{\dagger}}$, as given in Eq.~\ref{GfiniteTx} for
the case of finite $\lambda_1<0$ in the 2IK model.

Now we define a $4 \times 4$ unitary Fermi liquid scattering $S$
matrix for the 2IK model
\begin{equation}
\label{SMM}
S^{2IK}_{\sigma \alpha, \sigma' \alpha'}=-\left ( M^{>} \cdot
  {M^{<}}^{\dagger} \right )_{\sigma \alpha, \sigma' \alpha'}
\end{equation}
such that
\begin{equation}
\langle \psi_{\sigma \alpha}(z_1) \psi^\dagger_{\sigma' \alpha'}(z_2) \rangle_{H_{QCP}} = S_{\sigma \alpha, \sigma' \alpha'}  G_{\sigma \alpha, \sigma \alpha}\left( \frac{z_1 - z_2}{2i} \right).
\end{equation}
Using Eq.~\ref{M1} one obtains
$S_{\sigma \alpha, \sigma' \alpha'} =-\delta_{\sigma \sigma'}
(\delta_{\alpha \alpha'} \cos \theta - i \sin \theta \tau^x_{\alpha
  \alpha'} )=\delta_{\sigma \sigma'}  \frac{-\lambda_1 \delta_{\alpha
    \alpha'}+ i \lambda_2 \tau^x_{\alpha \alpha'}}{\lambda}$ for the
case of finite $\lambda_1$ and $\lambda_2$ in the 2IK model;
consistent with Eq.~\ref{s2ik}.

For arbitrary combination of $\{\lambda_1,...,\lambda_6 \}$, one has
\begin{equation}
\begin{split}
U&=e^{- \theta \int_{-\infty}^\infty dx~{\rm{sign(x)}} \chi_2^X [\lambda_2 \chi_1^f + \lambda_3 \chi_2^f + \vec{\lambda}_B  \cdot \vec{\chi}_s ] /\lambda_\perp} \\
 &= e^{i \frac{\theta}{2} \int_{-\infty}^\infty dx~{\rm{sign(x)}} \psi^\dagger [\lambda_2 \tau^x+ \lambda_3 \tau^y + (\vec{\lambda}_B  \cdot \vec{\sigma}) \tau^z ] \psi /\lambda_\perp}.
\end{split}
\end{equation}
From Eq.~\ref{M}, it then follows that
\begin{equation}
M^\gtrless = \cos \frac{\theta}{2} \mp i \sin \frac{\theta}{2} \left(\frac{\lambda_2 \tau^x+ \lambda_3 \tau^y + (\vec{\lambda}_B  \cdot \vec{\sigma}) \tau^z}{\lambda_\perp} \right),
\end{equation}
(suppressing spin and channel indices). Using Eq.~\ref{SMM} we recover our final
result for the 2IK model, Eq.~\ref{s2ik}.

Following the same steps for the 2CK model (and noting Eq.~\ref{2ik2ckcorr}), we obtain
\begin{equation}
S^{2CK}_{\sigma \alpha, \sigma' \alpha'}=-\left ( M^{>} \tau^z
  {M^{<}}^{\dagger} \right )_{\sigma \alpha, \sigma' \alpha'}
\end{equation}
with
\begin{equation}
M^\gtrless = \cos \frac{\theta}{2} -i \sin \frac{\theta}{2} \left
  (\frac{\lambda_2 \tau^y- \lambda_3 \tau^x \pm (\vec{\lambda}_B
    \cdot \vec{\sigma} )\tau^z}{\lambda_\perp}\right ),
\end{equation}
yielding precisely Eq.~\ref{s2ck} for the 2CK model.


\section{Numerical Renormalization Group}
\label{nrgcomp}
Wilson's numerical renormalization group\cite{wilson} (NRG) 
has been firmly established as a powerful technique for the accurate solution
of a wide range of quantum impurity problems.\cite{nrg:rev} 
Its original formulation provided access to numerically-exact
thermodynamic quantities for the Kondo\cite{wilson} and Anderson
impurity\cite{KWW} models. An increase in available computational resources
subsequently allowed straightforward extension to multi-impurity and
multi-channel systems.\cite{nrg:rev}

\begin{figure*}[t]
\begin{center}
\includegraphics*[width=110mm]{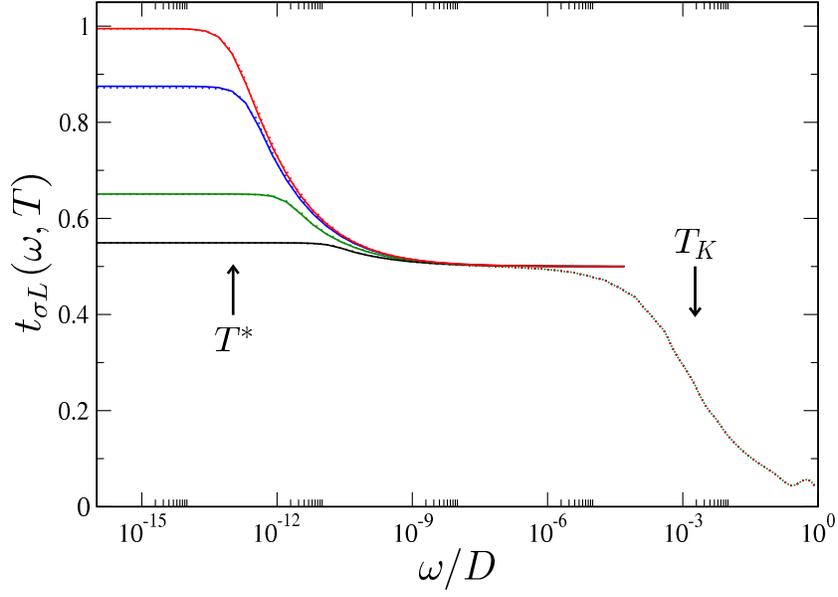}
\caption{\label{fig:finiteTnrg}
Spectrum $t_{\sigma L}(\omega,T)$ vs $\omega/D$ for the 2CK model with
$\nu J=0.15$ and small channel anisotropy $2\nu \Delta_z=10^{-6}$ at various
temperatures $T/T^*=10^{-1},1,10,10^2$, approaching $t_{\sigma
  L}=\tfrac{1}{2}$ from above. Dotted lines are results from full NRG
calculations; solid lines are exact results from Eq.~\ref{exactresult}
for the NFL to FL crossover. 
}
\end{center}
\end{figure*}

More recently, the identification of a complete basis within NRG (the
`Anders-Schiller basis' comprising discarded states across all
iterations\cite{asbasisprl}) has permitted rigorous extension to
calculation of dynamical quantities. In particular, equilibrium
spectral functions can be calculated using the full density matrix
approach,\cite{asbasis,fdmnrg} yielding essentially exact results at
zero-temperatures on all energy scales. Although discrete NRG data
must be broadened to produce the continuous spectrum,\cite{fdmnrg} artifacts
produced by such a procedure are effectively eliminated by averaging
over several interleaved calculations (the so-called
$z$-trick\cite{ztrick}). Indeed, resolution at high-energies can be
further improved by treating the hybridization term exactly.\cite{UFG}

Our exact analytic results were tested and confirmed by comparison to
NRG at $T=0$ in Ref.~\onlinecite{SelaMitchellFritz}.

Due to the logarithmic discretization of the conduction band inherent
to NRG,\cite{wilson} finite-temperature dynamical information cannot
however be captured\cite{fdmnrg}  on the lowest energy scales
$|\omega|\lesssim T$. But spectral functions for $|\omega|>T$ are
accurately calculated, 
and the total normalization of the spectrum is
guaranteed,\cite{fdmnrg} implying that the total \emph{weight}
contained in the spectrum for $|\omega|<T$ can be deduced. From a 
scaling perspective, one expects RG flow to be cut off on the energy
scale $|\omega|=\mathcal{O}(T)$, so that there should be no further
crossovers in spectral functions for $|\omega|<T$. The somewhat
arbitrary strategy\cite{fdmnrg} commonly employed is thus to smoothly
connect the spectrum calculated at $\omega\approx \pm T$, in such a
way as to preserve the total weight.

Our exact finite-temperature results for the crossover t matrix of the
2CK and 2IK models thus offers the perfect opportunity to benchmark
NRG calculations for interacting systems exhibiting a non-trivial
temperature-dependence of their dynamics. 

For concreteness, we consider now the 2CK model with
channel-anisotropy $\Delta_z>0$ (see Eqs.~\ref{2ck} and
\ref{delta2CK}). To obtain the numerical results, we discretize 
flat conduction bands of width $2D$ logarithmically using
$\Lambda=5$, and retain $8000$ states per iteration in each of $z=6$
interleaved NRG calculations.\cite{nrg:rev} All model symmetries are
exploited.

To ensure the desired scale separation $T^*\ll T_K$, we take
representative $\nu J=0.15$ and small $2\nu \Delta_z=10^{-6}$,
yielding $T_K/D= 2\times 10^{-3}$ and $T^*/D=9\times 10^{-14}$
($T_K\sim D \exp(-1/\nu J)$ is defined in practice here from the t
matrix Eq.~\ref{specdef},
$t_{\sigma\alpha}(\omega=T_K,T=0)=\tfrac{1}{4}$. $T^*$ is defined
according to Eq.~\ref{exactresult}, corresponding here to
$t_{\sigma\alpha}(\omega=T^*,T=0)\simeq 0.95$). From Eq.~\ref{tstar} and
Table~\ref{table:pert}, we thus obtain $c_1\approx 14$. The t matrix
for this 2CK model can be expressed as, 
\begin{equation}
\label{nrg_tmatrix}
\mathcal{T}_{\sigma\alpha,\sigma'\alpha'}(\omega,T) =\delta_{\sigma\sigma'}\delta_{\alpha\alpha'}
\left (\frac{J_{\alpha}}{2}\right )^2 \tilde{G}_{\alpha}(\omega,T)
\end{equation}
with $J_{\alpha}=J\pm \tfrac{1}{2}\Delta_z$ for $\alpha=L,R$ and where
\begin{equation}
\label{nrg_Gidef}
\tilde{G}_{\alpha}(\omega,T) = \langle\langle
\hat{S}^{-} \psi^{\phantom{\dagger}}_{0 \downarrow\alpha }+\hat{S}^z
\psi^{\phantom{\dagger}}_{0 \uparrow\alpha }; \hat{S}^{+}
\psi^{\dagger}_{0 \downarrow\alpha }+\hat{S}^z \psi^{\dagger}_{0 \uparrow\alpha } \rangle\rangle_{\omega,T}^{\phantom\dagger}.
\end{equation}
As usual $\langle\langle \hat{A};\hat{B}
\rangle\rangle_{\omega,T}^{\phantom\dagger}$ is the Fourier transform of
the retarded correlator $\langle\langle \hat{A}(t_1);\hat{B}(t_2)
\rangle\rangle_T=-i\theta(t_1-t_2)\langle\{\hat{A}(t_1),\hat{B}(t_2)\}\rangle_T$.
 The alternative expression given in Ref.~\onlinecite{akm:oddimp} is:
\begin{equation}
\label{nrg_tmatrix2}
\pi\nu \mathcal{T}_{\sigma\alpha,\sigma\alpha}(\omega,T) =
-i\left [1+\left ( \frac{2}{\pi\nu J_{\alpha}} \right )^2 \frac{G_{\alpha
}(\omega,T)}{\tilde{G}_{\alpha}(\omega,T)} \right ]^{-1},
\end{equation}
where $G_{\alpha}(\omega,T) = \langle\langle
\psi^{\phantom{\dagger}}_{0 \sigma \alpha };\psi^{\dagger}_{0 \sigma\alpha}
\rangle\rangle_{\omega,T}^{\phantom\dagger}$ is the Green function for
the `0'-orbital of the $\alpha=L,R$ Wilson chain.\cite{nrg:rev} 
Both $\tilde{G}_{\alpha}(\omega,T)$ and $G_{\alpha}(\omega,T)$ can be
obtained directly by NRG, but Eq.~\ref{nrg_tmatrix2} gives much better
numerical accuracy,\cite{akm:oddimp} and is employed in the following. 
The desired spectral function $t_{\sigma\alpha}(\omega,T)$ is then
obtained from Eq.~\ref{specdef}, and is plotted in
Fig.~\ref{fig:finiteTnrg} as the dotted lines for temperatures
$T/T^*=10^{-1},1,10,10^2$, as in Fig.~\ref{fig:k}. The corresponding
exact results for the NFL to FL crossover from Eq.~\ref{exactresult}
are plotted as the solid lines. As immediately seen, near-perfect
agreement is obtained for all energies
$|\omega|\ll T_K$ and temperatures $T\ll T_K$ where comparison between
numerical and exact results can be made.

To obtain such an agreement, we found that high-accuracy NRG 
calculations must be performed. In particular, the region
$|\omega|\sim T$ was most problematical, with artifacts only being
removed upon averaging over several band discretizations, and
necessitating a large number of states to be kept at each NRG
iteration.  The precise shape of the numerically-obtained spectrum
then still depends on how the discrete data is smoothed. We found that the
broadening scheme described in Ref.~\onlinecite{fdmnrg} produced the best
results: for $z=6$ and $\Lambda=5$ as used here, a broadening
parameter $b=0.25$ and kernel-crossover scale $\omega_0=T/1.5$
were optimal. 

It should also be noted that if the correction factor\cite{nrg:rev} 
$A_{\Lambda}=\tfrac{1}{2}\tfrac{\Lambda+1}{\Lambda-1}\log(\Lambda)$ is
used in the NRG calculations (in which case $J_{\alpha}\rightarrow
J_{\alpha} A_{\Lambda}$), then the many-particle energies used to calculate the
density matrix must be accordingly scaled ($E_N(r)\rightarrow
E_N(r) A_{\Lambda}$) so that the results are independent of the
discretization parameter $\Lambda$ and hence approximate accurately
the desired $\Lambda=1$ limit.


\section{Other exact crossover functions}
\label{other}
As discussed in the previous sections, the Hamiltonian controlling the NFL--FL
crossover in the 2CK or 2IK models has a free fermion structure in
terms of MFs. In fact, this feature allows calculation of various
quantities along the crossover. The difficulty of such calculations is dictated by
the relation between the physical quantity of interest to the
MFs. In the preceding sections we concentrated on the two-point
function of the electron field (the Green function),
related\cite{cardy} here to the one-point function of the
magnetization operator in the Ising model (which is in
turn related non-locally to the Ising MFs\cite{zuber}).
More generally, $2p$-point functions of the electron field are
related\cite{cardy} to $p$-point functions of the magnetization
operator. Multi-electron correlators can thus in principle be calculated
in this way, but require knowledge of the corresponding multi-point correlation
functions of the Ising magnetization operator.

\begin{figure*}[t]
\begin{center}
\includegraphics*[width=145mm]{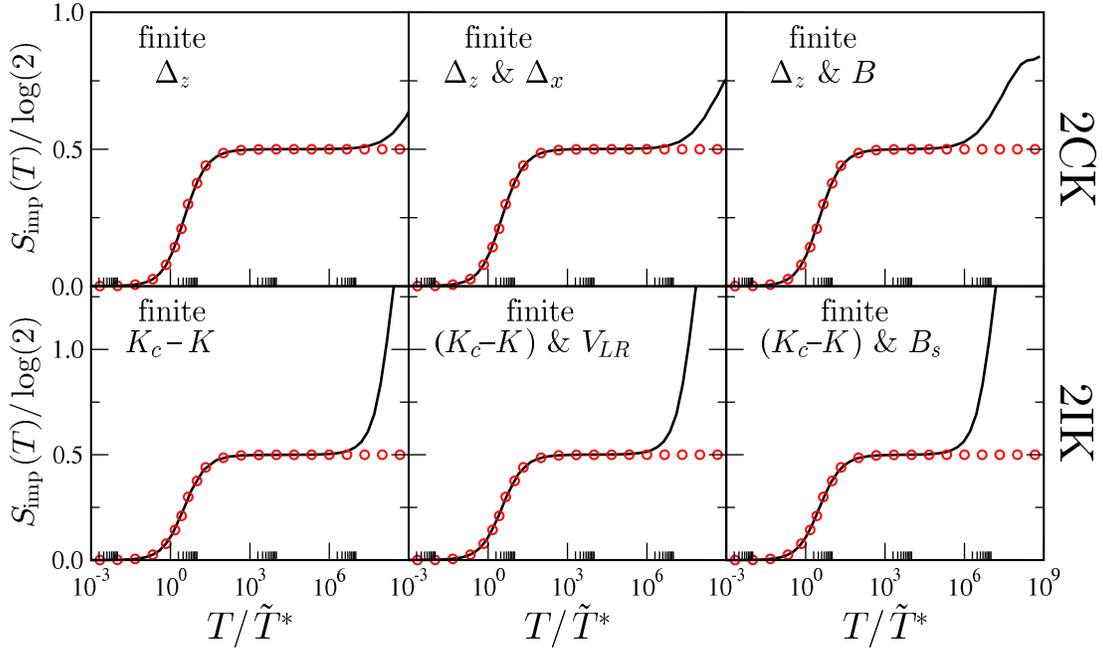}
\caption{\label{fig:entcross}
Impurity contribution to entropy $S_{\text{imp}}(T)$ vs
$T/\tilde{T}^*$ for the 2CK model (upper panels) and the 2IK model
(lower panels) in the presence of various perturbations. Entire
temperature-dependence calculated by NRG for the full models (black
lines); low-temperature $T\ll T_K$ behavior in each case compared
with the single exact NFL--FL crossover function Eq.~\ref{entropycross} (red
circles). All results presented for $\nu J=0.25$.
\emph{Left}: effect of channel asymmetry $4\nu\Delta_z = \pm 10^{-5}$
(2CK) or deviation from critical coupling $(K_c-K)/D=\pm 10^{-5}$ (2IK).
\emph{Center}: effect of including also finite left/right tunneling,
$2\nu\Delta_x = \pm 10^{-5}$ (2CK) or $2\nu V_{LR} = \pm 10^{-4} $
(2IK), with $4\nu\Delta_z = (K_c-K)/D=\pm 10^{-5}$ as before.
\emph{Right}: effect of including finite magnetic field,
$B^z/D=\pm 10^{-9/2}$ (2CK) or $B_s^z/D=\pm 10^{-9/2}$ (2IK), again with
$4\nu\Delta_z = (K_c-K)/D=\pm 10^{-5}$. Parameters chosen to allow
direct comparison to Fig.~1 of Ref.~\onlinecite{SelaMitchellFritz}.
}
\end{center}
\end{figure*}

\subsection{Impurity entropy}
Since thermodynamic quantities are local in the MFs, their calculation
is rather straightforward. Here we will focus on the NFL--FL crossover
of the impurity entropy, following closely the earlier calculations
for the 2CK model\cite{gogolin,emery} and the 2IK model,\cite{gan}
performed in the Toulouse limit. The Toulouse limit corresponds here
to maximal spin anisotropy in the exchange couplings, and as such
breaks the overall $SU(2)$ spin symmetry of the models. Although the
high-energy $\sim T_K$ crossover to the NFL fixed point is
strongly affected by large spin-anisotropy, we
stress that for low energies $\ll T_K$ (and given a clear scale
separation $T^* \ll T_K$), the results become formally exact, and are
universally applicable to the $SU(2)$ symmetric case of interest.

The key point is that the spin-anisotropy perturbation is RG
\emph{irrelevant} at the NFL fixed point. In particular, the effective
theory obtained in the Toulouse limit describing the NFL--FL crossover
due to relevant perturbations such as channel anisotropy or magnetic
field in the 2CK model\cite{gogolin} or detuning $K-K_c$, staggered
magnetic field, or left-right tunneling in the 2IK model,\cite{gan}
act exactly as in Eq.~\ref{deltaH}. A detailed discussion for the 2IK
model can be found in Ref.~\onlinecite{SelaPRL}.

Turning now to the crossover in the impurity entropy, one finds\cite{gogolin} that
\begin{equation}
\label{entropycross}
S(T) = \tfrac{1}{2}\log (2) +\bar{S}\left( \frac{T}{\tilde{T}^*} \right),
\end{equation}
in terms of the universal function
\begin{equation}
\bar{S}(t) = \frac{1}{t} \left[ \psi\left(
    \frac{1}{2}+\frac{1}{ t} \right)-1\right ]- \log \left
  [ \frac{1}{\sqrt{\pi}}\Gamma \left( \frac{1}{2}+\frac{1}{ t}
  \right)\right ],
\end{equation}
defined in Ref.~\onlinecite{gogolin} for the limit $T_K \to
\infty$. Here, $\psi(z)$ is the psi (digamma) function and
$\tilde{T}^*$ is a particular definition of the NFL--FL crossover
scale (proportional to our definition, Eq.~\ref{exactresult}, such that $\tilde{T}^*=y
\times T^*$ with $y\approx 4.6$).
Two regimes can thus be distinguished. In the FL regime, obtained for
$T \ll T^*$, the impurity is always completely screened: $S \sim
\frac{1}{12} \left(\frac{T}{\tilde{T}^{*}} \right)$. By contrast, in the NFL
regime, $T^* \ll T \ll T_K$, the impurity entropy is close to
$\tfrac{1}{2}\log (2)$. Interestingly, we find that independently of the
relevant perturbations which act, the entropy crossover is
always given by the universal function Eq.~\ref{entropycross} in the limit $T^*
\ll T_K$, in both 2CK and 2IK models. This is of course not the case
for the Green function, because the FL scattering S matrix is affected
differently by different perturbations (see
Eqs.~\ref{exactresult}--\ref{analyticformula} and Figs.~\ref{fig:k},
\ref{fig:mag}).

The 2CK model has also been solved exactly using the Bethe
ansatz,\cite{s} yielding
the full evolution of thermodynamics in any parameter regime. However,
it cannot be seen directly from the Bethe ansatz equations
that there is an emergent $SO(8)$ symmetry at the NFL fixed point, or
that this leads to a single NFL--FL crossover function for the
entropy, Eq.~\ref{entropycross}, regardless of the perturbation
causing the crossover. Indeed, the fact that the same crossover occurs
in the 2IK model cannot be extracted using Bethe ansatz since the 2IK
model is not integrable.

In Fig.~\ref{fig:entcross} we present numerically-exact NRG results for the
temperature-dependence of the entropy due to various perturbations in
the 2CK and 2IK models to confirm the validity of the field theoretic
description. 

As in Fig.~\ref{fig:finiteTnrg}, we exploit all model symmetries to
obtain high-quality numerics, discretizing flat 
conduction bands of width $2D$ logarithmically, using $\Lambda=3$
here, and retaining $8000$ states per iteration in a single NRG
calculation.\cite{nrg:rev} 

At low temperatures $T\ll T_K$ (and since $T^*\ll T_K$),
we obtain an essentially perfect agreement between the exact result
Eq.~\ref{entropycross} (points) and NRG data (solid line).

\subsection{Non-equilibrium transport in two lead devices}
It was shown in Refs.~\onlinecite{SelaPRL,Sela2009PRB,SelaPRL2} that
the effective free fermion theory of the 2IK model allows to calculate
certain \emph{non-equilibrium} quantities. Finite conductance was
found to arise in the weak coupling limit of 2IK systems close to the
critical point $T^*\ll T_K$ at low energies $\ll T^*$. This result was
understood in terms of the growth under RG of the left-right tunneling
perturbation $V_{LR}$.  Here, we generalize these results to the 2CK
model, which has the same effective free fermion description. Related
multichannel setups have been considered in Refs.~\onlinecite{mitra,schuricht}.

\begin{figure}[b]
\begin{center}
\includegraphics*[width=40mm]{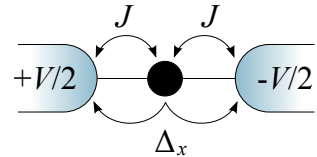}
\caption{\label{fig:dot} Schematic illustration of a non-equilibrium
  2CK setup. We consider the case of $\Delta_x\ll J$.}
\end{center}
\end{figure}
We consider a finite source-drain voltage $V$ across left and right metallic
leads, which are exchange-coupled to a single impurity
spin. To this system we add a small but finite channel anisotropy
perturbation, corresponding left/right tunneling mediated
via the impurity spin. The setup is illustrated in
Fig.~\ref{fig:dot}. The corresponding Hamiltonian is given by
Eqs.~\ref{2ck} and \ref{delta2CK}, with finite $\Delta_x$ and possibly
magnetic field $\vec{B}$, but now with left/right lead chemical potentials at
$\pm V/2$.

The applicability of our exact solution is in the parameter regime
$\Delta_x\ll J$, so that the system is close to the NFL critical
point. This situation is not in practice obtained in standard quantum
dot devices, although more sophisticated experimental techniques such
as those employed in Ref.~\onlinecite{Potok07}, do allow suppression
of cotunneling perturbations such as $\Delta_x$.

As per Eq.~\ref{tstar}, the crossover energy scale is $T^*=
(c_1\nu \Delta_x)^2 T_K +|c_B B|^2/T_K$. In the limit where $\nu \Delta_x$ is
initially very small, we thus have  $T^* \ll T_K$. At higher energies
$\gtrsim T_K$, we then expect conductance to be very small $\propto
(\nu \Delta_x)^2$, corresponding to the weak coupling limit. However,
upon reducing the energy scale $E = \max \{ eV , T \}$  below $T_K$,
the conductance starts to increase since $\Delta_x$ switches on a
relevant operator with scaling dimension $1/2$ near the NFL FP.
Below $T^*$, a characteristic peak in the conductance is thus
expected, signaling growth of the relevant operator to order one.

\begin{figure}[t]
\begin{center}
\includegraphics*[width=70mm]{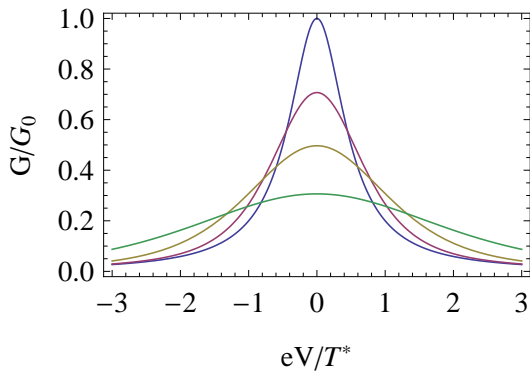}
\caption{\label{fig:cond} Scaling function for the nonlinear
  conductance of a 2CK device with small $\Delta_x$. From top to
  bottom at the peak: $T/T^*=0,0.25,0.5,1$.} 
\end{center}
\end{figure}

The exact lineshape of the non-equilibrium conductance peak can be
calculated from the fixed point Hamiltonian, Eq.~\ref{HQCP} (including
the correction due to relevant perturbations given by Eq.~\ref{deltaH}). The
dependence on the ratio between the magnetic field and the tunneling
perturbations is obtained using the $SO(8)$ rotation outlined in
Sec.~\ref{gengf}. The method of calculation, and the result for the
universal lineshape of the peak, was obtained for the spin-exchange anisotropic
version of the model by Schiller and Hershfield in Ref.~\onlinecite{shiller}. As
argued in the previous subsection, we can borrow those Toulouse-limit
results if we restrict attention to the low energy crossover. The
final result for the nonlinear conductance is thus
\begin{equation}
\begin{split}
G = & G_0 F\left[ \frac{T}{T^*} ,\frac{eV}{T^*} \right],\\
F[t,v] = &\frac{1}{4 \pi t} {\rm{Re}}~ \psi_1 \left(
  \frac{1}{2}+\frac{1}{4 \pi t}+\frac{i v}{2 \pi t}\right),
\end{split}
\end{equation}
with $G_0=\frac{2e^2}{h} \frac{\lambda_2^2+\lambda_3^2}{\lambda^2}$ and $\psi_1$ the
trigamma function. Note that the definition of $T^*$ here is as in
Ref.~\onlinecite{SelaPRL}. At $T=0$, one
obtains $G/G_0 = [1+(2 eV / T^*)^2]^{-1}$; while at zero-bias $V=0$
and low-temperatures $T \ll T^*$, the asymptotic conductance is
$G/G_0 \to 1 - (2 \pi T / \sqrt{3} T^*)^2$. The full bias-dependence
of conductance for various temperatures is shown in
Fig.~\ref{fig:cond}.


\section{Conclusions}
\label{concs}
In this paper we present a rare example of an exact nonperturbative
result for the finite-temperature dynamics of a strongly correlated
quantum many-body system. 
We focus on the two-channel Kondo and two-impurity Kondo models;
although the same low-energy physics characterizes a wide class of 
quantum impurity problems in which competition between two conduction
channels causes a frustration of screening. 
The unusual non-Fermi liquid critical points of these systems are
destabilized by various symmetry-breaking perturbations, naturally
present in experiment. In consequence, a crossover to regular Fermi
liquid behavior always occurs on the lowest energy scales.    
Exploiting the connection\cite{CFT2IKM} to an
exactly-solved classical boundary Ising model,\cite{cz,Leclair} we
calculated the exact finite-temperature crossover Green function. 
In quantum dot systems which could access this crossover, the relevant
experimental quantity is conductance, which we extract from the exact
Green function. 
 
Remarkably, we show that due to the free fermion structure of the
effective low-energy theory in terms of Majorana fermions and a large
emergent $SO(8)$ symmetry, a \emph{single universal function} pertains for
any combination of perturbations in either model. This single
crossover is also starkly manifest in the behavior of thermodynamic
quantities such as entropy; as confirmed directly by NRG. 

The method developed in this paper goes beyond the impurity models 
we considered explicitly, and finds powerful application to a wider
family of systems. At heart, our solution relies upon a formal
separation of the theory into a sector containing all the universal
crossover physics, and a sector acting as a spectator along 
this crossover. 
Importantly, the crossover is confined to a sector which can
be identified with Ising degrees of freedom, described by a 
minimal conformal field theory with central charge $c=1/2$. 
For example, in the two-channel Kondo models studied here, the
full set of degrees of freedom consist of a $c=4$ CFT, but a large $c=7/2$
sector of the theory plays no role in the crossover from non-Fermi
liquid to Fermi liquid physics.  

Interestingly there exist other models (whose full set of
degrees of freedom are not necessarily described by a $c=4$ CFT) which
undergo precisely the same crossover due to their underlying $c=1/2$
Ising sector. Those include certain Luttinger liquids containing an
impurity,\cite{Leclair} and coupled bulk 
and edge states in certain non-Abelian fractional quantum Hall 
states\cite{rosennow,Bishara} (see also Ref.~\onlinecite{Sevier}).

There are further interesting generalizations and questions arising
from this work. For example, the two-channel Kondo effect evolves
continuously as interactions are switched on in the leads, as was
shown in the case of Luttinger liquid\cite{LL_fab,LL_chandra} 
and helical liquid\cite{HL_schiller,HL_oreg,HL_law} leads. 
It is an open question as to whether the low-temperature crossovers in the
presence of such interacting leads are described by the same boundary
Ising model, or e.g.~by coupled boundary Ising models. It would also be
interesting to use the present formulation of the crossover in terms
of a minimal Ising theory to study time-dependent phenomena, quench
dynamics, and other non-equilibrium physics.


\begin{acknowledgments}
We thank L. Fritz, H. Saleur and A. Rosch for helpful
discussions. This work was supported by the A.~v.~Humboldt Foundation
(E.S.) and by the DFG through SFB608 and FOR960 (A.K.M.).
\end{acknowledgments}


\appendix


\section{Perturbation theory around the FL fixed point}
\label{pertFL}
In this appendix we consider the 2IK model for $K<K_c$, and its FL
fixed point describing the ground state where each impurity forms
a Kondo singlet with its attached lead. In particular, we calculate the t
matrix for $\omega, T \ll T^*$ as a stringent consistency check of our
full crossover t matrix, Eq.~\ref{exactresult}. Indeed, we also see that
the multiplicative function $f(\beta h^2)$ that we included in
Eq.~\ref{finitetisingmag} is precisely needed to reproduce the correct FL
limit.

We use here the Fermi liquid theory of Nozi\`{e}res,\cite{Nozieres} applied to the 2IK
model. Naturally, our derivation of the t matrix in the vicinity of the
FL fixed point follows closely the analogous calculation for the
simpler single channel Kondo (1CK) model. Thus, we first recap some of the
basic concepts and results for the 1CK model.\cite{Nozieres,CFT2CK,Selemalecki}

\begin{figure}[h]
\begin{center} \includegraphics*[width=45mm]{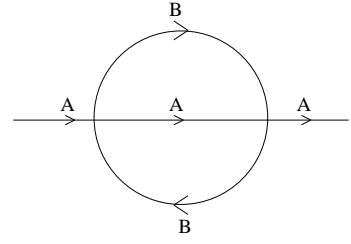}
    \caption{Diagram for the inelastic contribution to the t matrix. It describes interaction between  two fermionic species $A \ne B$. In the 1CK model $A,B \in \{\uparrow, \downarrow \}$; in two-channel Kondo models $ A,B  \in \{\uparrow L, \downarrow L, \uparrow R , \downarrow R \}$.
\label{fig:fldiag}}
\end{center}
\end{figure}

The irrelevant operator in the effective Nozi\`{e}res Hamiltonian\cite{Nozieres} for
the FL fixed point of the 1CK problem may be written in CFT language as\cite{CFT2CK}
\begin{equation}
\label{nozieres}
\begin{split}
\delta H_{1CK} &= - \frac{1}{T_K} \vec{J} (0) \cdot \vec{J} (0) \\
&= -   \frac{3}{2T_K}({\psi'}^\dagger_\sigma i \partial_x {\psi'}_\sigma - {\psi'}^\dagger_\uparrow {\psi'}_\uparrow {\psi'}^\dagger_\downarrow {\psi'}_\downarrow) _{x=0},
\end{split}
\end{equation}
where $\vec{J}(x) = {\psi'}^\dagger_{ \sigma}(x) \frac{\vec{\sigma}_{\sigma
    \sigma'}}{2} {\psi'}_{ \sigma'} (x)$ is the spin current for a single
channel of conduction electrons (implicit summation over repeated
indices is implied). The first term of the second line may be
interpreted as an  elastic single particle scattering, and the second
term can be interpreted as a residual electron-electron interaction
giving rise to inelastic scattering. Accordingly, one can  separate
the contributions to the t matrix as
$\mathcal{T}_{1CK}\left( w,t \right) = \mathcal{T}_{el}\left( w,t
\right)+\mathcal{T}_{in}\left( w,t \right)$, where\cite{CFT2CK,Selemalecki}
\begin{equation}
\label{tmatrixfl}
\begin{split}
- \pi \nu \mathcal{T}_{el}\left( w,t \right) &= i -w - i w^2, \\
 - \pi \nu \mathcal{T}_{in}\left( w,t \right) &=  -\frac{i}{2} [w^2+ \pi^2 t^2 ].
\end{split}
\end{equation}
[For simplicity we omit spin indices; $\mathcal{T}_{1CK \sigma \sigma'}(w,t) = \delta_{\sigma \sigma} \mathcal{T}_{1CK}\left( w,t \right)$]. Here $w=\omega/T_K'$, $t=T/T_K'$, and $T_K'$ is a particular
definition of the Kondo temperature.\cite{CFT2CK} The fermionic
diagram yielding the inelastic contribution is shown in
Fig.~\ref{fig:fldiag}.

The imaginary part of the t matrix in the 1CK model can thus be expanded as
\begin{equation}
\text{Im}~\mathcal{T}_{1CK}\left( w,t \right)
=\text{Im}~\mathcal{T}_{1CK}\left( 0,0 \right) + a w^2+b t^2,
\end{equation}
where from Eq.~\ref{tmatrixfl} one obtains
\begin{equation}
a/b|_{1CK} =3/\pi^2\simeq 0.3039.
\end{equation}

Similar to the Nozi\`{e}res Fermi liquid theory for the 1CK model, an
effective Fermi liquid Hamiltonian was constructed using CFT methods based on the emergent SO(7) symmetry of the crossover
for the 2IK model.\cite{SelaPRL2,MaleckiSelaAffleck} The leading irrelevant operator
takes the form\cite{SelaPRL2}
\begin{equation}
\label{FL}
\delta H_{FL}=\frac{1}{T^*} ( \vec{J}_L^2+\vec{J}_R^2  - 6  \vec{J}_L \vec{J}_R )_{x=0},
\end{equation}
where $\vec{J}_\alpha (x)= {\psi'}^\dagger_{\alpha \sigma}(x)
\frac{\vec{\sigma}_{\sigma \sigma'}}{2} {\psi'}_{\alpha \sigma'} (x)$,
${\psi'}$ is a scattering state incorporating the $\pi/2$ Kondo phase
shift, ${\psi'}_{\alpha \sigma}(x) = {\rm{sign}}(x)\psi_{\alpha \sigma}(x)
$, and $T^*$ is a particular definition of the low
energy crossover scale.\cite{SelaPRL2}
We now calculate the t matrix resulting from this Hamiltonian in the
FL regime $\omega, T \ll T^*$. In the following we suppress the
indices $\sigma,\alpha$, and note that the t matrix is proportional to
$\delta_{\sigma \sigma'} \delta_{\alpha \alpha'}$ in the present
situation.

Comparison of the irrelevant operators in the 1CK and 2IK models,
Eqs.~\ref{nozieres} and \ref{FL}, shows that the first two terms in
Eq.~\ref{FL} are identical to the Nozi\`{e}res irrelevant operator (up
to the exchange of energy scales $T^* \leftrightarrow T_K$). In
consequence the elastic and inelastic scattering contributions
\emph{within} each channel of the 2IK model are the same as those
arising in the 1CK model. Indeed, from the diagram
Fig.~\ref{fig:fldiag} we see that to second order they do not yield
any mixed terms. Thus $\mathcal{T}_{2IK}\left( w,t \right) =
\mathcal{T}_{1CK}\left( w,t \right)+ \mathcal{T}_{LR}\left( w,t
\right) $, where $ \mathcal{T}_{LR}\left( w,t \right) $  originates
from the third term in Eq.~\ref{FL}, representing interaction
\emph{between} channels. We separate the latter into $\vec{J}_L \cdot
\vec{J}_R =\sum_{a=x,y,z} J_L^a J_R^a $ and note that the second order
$a=x,y,z$ contributions are equal, since the 
quantity of interest is invariant under spin rotations, and the
Hamiltonian is also $SU(2)$ spin-symmetric. The latter can be written
in terms of fermion fields as
\begin{equation}
\label{flint}
-\frac{6}{T^*} J_L^z J_R^z =-\frac{3}{2 T^*} ({\psi'}_{\sigma L }^\dagger \sigma^z_{\sigma \sigma} {\psi'}_{\sigma L }) ({\psi'}_{\sigma R }^\dagger \sigma^z_{\sigma \sigma} {\psi'}_{\sigma R }).
\end{equation}

Considering now the t matrix for a single electronic species with quantum numbers
$A=\sigma,\alpha$, Eq.~\ref{flint} describes the interaction with a
second species of either $B = \sigma,\alpha$ or $\bar{\sigma},\alpha$,
and thus contributes a term proportional to the inelastic contribution
$\mathcal{T}_{in}\left( w,t \right)$.  In fact, the amplitude for this
interaction is $ \pm \frac{3}{2 T^*}$, identical in absolute value to
the intra-lead interaction amplitude between up and down electrons
[the second term of the second line in Eq.~\ref{nozieres}]. But since
the contributions to inelastic scattering are of second order (see
Fig.~\ref{fig:fldiag}), the \emph{sign} of the scattering amplitude is
unimportant in calculation of the t matrix. Summing over the second
species $B$ yields an extra factor of 2, yielding $
\mathcal{T}_{LR}\left( w,t \right)=3 \mathcal{T}^z_{LR}
= 6 \mathcal{T}_{in}\left( w,t \right)$. Putting all the contributions
together, we have
\begin{equation}
\mathcal{T}_{2IK}\left( w,t \right)  = \mathcal{T}_{el}\left( w,t \right)+7 \mathcal{T}_{in}\left( w,t \right).
\end{equation}
 As a result, one again obtains $\text{Im}~\mathcal{T}_{2IK}\left( w,t \right)
=\text{Im}~\mathcal{T}_{2IK}\left( 0,0 \right) + a w^2+b t^2$, but with
\begin{equation}
\label{abratio}
a/b|_{2IK} = (1+2/7)/\pi^2 \simeq 0.13027.
\end{equation}
This result is in perfect agreement with a numerical evaluation of our
full finite-temperature crossover t matrix, Eq.~\ref{exactresult} in
the limit $T,\omega \ll T^*$; as demonstrated in
Fig.~\ref{fig:tw}. This calculation also confirms the need for the
function $f(\beta h^2)$ used in Eq.~\ref{finitetisingmag}.

The asymptotic result Eq.~\ref{abratio} also follows from renormalized
perturbation theory calculations presented recently in 
Ref.~\onlinecite{hewsonRPT} for a related 2IK model.


\section{Perturbation theory around the NFL fixed point}
\subsection{General structure}
\label{app_analyticity}
In Sec.~\ref{gft0} we used the analyticity of the Green function and
the locality of the t matrix to argue that
$G_{\sigma\alpha,\sigma'\alpha'}(z_1-z_2)$ depends only on the
difference $(z_1-z_2)$. This result should hold to all orders in
perturbation theory, as shown explicitly in this appendix.

Our starting point here is the NFL FP Green function for the 2IK
model, Eq.~\ref{factorG}, written as,
\begin{equation}
\label{Gnfl}
\begin{split}
G_{\sigma\alpha,\sigma'\alpha'}^{\textit{NFL}}(z_1,z_2)=\delta_{\sigma\sigma'}\delta_{\alpha\alpha'} \left
  (\frac{\tfrac{1}{2\pi}\tfrac{\pi}{\beta}}{\sin[\tfrac{\pi}{\beta}(z_1-z_2)]} \right
)^{\tfrac{7}{8}} \\
\times \langle \sigma_L(z_1)\sigma_L(z_2) \rangle,
\end{split}
\end{equation}
whose factorized form originates from the Bose-Ising decomposition of
the 2IK model into spin, isospin and Ising symmetry
sectors.\cite{CFT2IKM}
7 of the 8 MFs (those corresponding to the part of the fermion field
carrying spin and isospin quantum numbers) remain free at the NFL FP,
and thus give rise to the first factor (see Eq.~\ref{freeG}). The second
correlator involves the chiral part $\sigma_L(z)$ of the Ising magnetization
operator $\sigma(z)=\sigma_L(z)\sigma_L(z^*)$, due
to the remaining MF in the Ising sector (see Fig.~\ref{fig:images}).

When the detuning perturbation $\lambda_1\propto (K-K_c)$ acts, the
NFL FP is destabilized. The perturbation appears
as a correction to the action,\cite{ghosal}
\begin{equation}
\label{deltaS}
\delta S = \lambda_1 \int d \tau \epsilon(0,\tau).
\end{equation}
where $\epsilon$ is the CFT $d=1/2$ boundary operator from the Ising
sector interpreted as a boundary magnetic field.\cite{CFT2IKM}
The resulting corrections to the Green function can then be calculated
within perturbation theory. The full crossover Green function is then 
$G_{\sigma\alpha,\sigma'\alpha'}^{\textit{NFL-FL}}(z_1,z_2)=G_{\sigma\alpha,\sigma'\alpha'}^{\textit{NFL}}(z_1-z_2)
+
\sum_{N=1}^{\infty}\delta_N
G_{\sigma\alpha,\sigma'\alpha'}(z_1,z_2)$, where the $N$-th order correction is given by (suppressing spin and channel indices)
\begin{equation}
\label{integral2}
\begin{split}
\delta_N G(z_1,z_2) \propto  \lambda_1^N &\left(
  \frac{\tfrac{1}{2\pi}\tfrac{\pi}{\beta}}{\sin[\frac{\pi}{\beta}
    (z_1 - z_2)]}
\right)^{\tfrac{7}{8}}  \int_0^\beta \prod_{i=1}^N d \tau_i  \times \\
 &\langle \sigma_L(z_1) \sigma_L (z_2)  \prod_{j=1}^N \epsilon(0,\tau_j)  \rangle.
\end{split}
\end{equation}

Generically, correlation functions up to 3-point functions are
determined by CFT. However, the Ising CFT is special because
essentially \emph{all} correlation functions are known exactly. In
particular, Ardonne and Sierra obtained explicit expressions\cite{ardonne} for the
correlators appearing in Eq.~\ref{integral2}. In the case of even $N$
their result reads,\cite{ardonne}
\begin{equation}
\label{eqardonne}
\begin{split}
 \langle \sigma_L(z_1) \sigma_L (z_2)  \prod_{j=1}^N \epsilon(0,\tau_j)  \rangle \propto (z_1 -z_{2})^{-1/8}  \times \\
 \sqrt{\sum_{I} 2^{|I|} \left( {\rm{Hf}}_{i,j\in I} \frac{1}{(\tau_i - \tau_j)^2} \right) \frac{(z_1 - z_2)^{|\tilde{I}|}}{\prod_{j \in \tilde{I}}(z_1-\tau_j)(z_2 - \tau_j)} },
\end{split}
\end{equation}
where the sum is over all subsets of $\{1,2,...,N \}$, containing an
even number of elements $|I|$. ${\rm{Hf}}(M)$ denotes the Haffnian of
a symmetric $N \times N$ matrix, and is given by
${\rm{Hf}}(M)=\frac{1}{2^{N/2}(N/2)!} \sum_{\sigma \in S_N}
\prod_{i=1}^{N/2} M_{\sigma(2i-1),\sigma(2i)}$, with $\sigma$  a
permutation. The set $\tilde{I}$ (containing $|\tilde{I}|$ elements)
is equal to $\{1,2,...,N \} \setminus I$.  Using the conformal mapping
Eq.~(\ref{confmap}) from the plane to the cylinder, each coordinate
difference $z-z'$ in Eq.~(\ref{eqardonne}) is replaced by
$\tfrac{\beta}{\pi} \sin
  \left[ \frac{\pi}{\beta} (z-z') \right]$.  The
dependence on $z_1$ and $z_2$ is through factors which explicitly
depend on $z_1 - z_2$, and terms inside the square-root of the form
\begin{equation}
\frac{1}{\prod_{j \in \tilde{I}} \sin \left( \frac{\pi}{\beta}(z_1 - \tau_j)\right) \sin \left( \frac{\pi}{\beta}(z_2 - \tau_j)\right)}.
\end{equation}
With the aid of the trigonometric identity,
\begin{equation}
\begin{split}
2&\sin\left ( \tfrac{\pi}{\beta}(z_1-\tau)\right ) \sin\left (
  \tfrac{\pi}{\beta}(z_2-\tau)\right ) \\ = & \cos\left (
  \tfrac{\pi}{\beta}(z_1-z_2)\right ) -  \cos\left (
  \tfrac{\pi}{\beta}(z_1+z_2-2\tau)\right ),
\end{split}
\end{equation}
and by shifting all $\tau_j$ integration variables by
$(z_1+z_2)/2$ into the complex plane (which can be done without
encountering any singularities), the resulting integral
in Eq.~\ref{integral2} then depends only on $z_1-z_2$.
The same conclusion is reached for odd-$N$ by a similar calculation.

Thus $\delta_N G_{\sigma\alpha,\sigma'\alpha'}(z_1,z_2)\equiv \delta_N
G_{\sigma\alpha,\sigma'\alpha'}(z_1-z_2)$ for all $N$, and hence the
full crossover Green function depends only on $(z_1-z_2)$. Analyticity
of the Green function
$\langle \psi_{\alpha \sigma}(x) \psi_{\alpha' \sigma'}^\dagger(-x)
\rangle \equiv G_{\sigma\alpha,\sigma'\alpha'}(x)$ thus allows determination of
$\langle \psi_{\alpha \sigma}(z_1) \psi_{\alpha'
  \sigma'}^\dagger(z_2)\rangle \equiv G_{\sigma\alpha,\sigma'\alpha'}\left (\frac{z_1-z_2}{2i} \right )$
in terms of general coordinates $z_1$ and $z_2$ by analytic
continuation.

\subsection{Leading order perturbation theory}
\label{NFLpert}
Here we derive the NFL coefficients $\beta',\delta$ and $\beta''$ of
the asymptotic t matrix discussed in Sec.\ref{results}. Since the Green function
vanishes at the NFL fixed point, the leading correction arises to
first order. We now use the first order result for $\langle 
\sigma(x) \rangle$ derived in Ref.~\onlinecite{ising_note}, 
\begin{equation}
\label{pt}
\begin{split}
\langle \sigma(x) \rangle_{ \beta}^{(1)} =& h \sqrt{2 \pi \beta }
\left( \frac{\frac{4 \pi}{\beta}}{\sinh  \frac{2 \pi x}{\beta}}
\right)^{1/8}\times \\
&_{2}F_1\left(\frac{1}{2},\frac{1}{2};1,\frac{1-\coth \frac{2 \pi
      x}{\beta}}{2} \right)+ \mathcal{O}(h^2).
\end{split}
\end{equation}
Note that $_{2}F_1\left(\frac{1}{2},\frac{1}{2};1,z \right) = \frac{2 K[z]}{\pi} $ with $K$ the complete elliptic integral of the first kind.
We will also use the short distance $x \to 0 $ limit of this formula,
\begin{equation}
\label{shortdT}
\langle \sigma(x) \rangle_{h ,\beta}  = -2^{13/8}h x^{3/8} [\log (x) +
\mathcal{O}(1)],
\end{equation}
valid for $x \ll \beta, h^{-2}$. Eqs.~\ref{factorGising} and
\ref{shortdT} give 
 \begin{equation}
 \label{Gsmallx}
 G(x \to 0) = -\frac{h }{\pi \sqrt{2} i } \frac{1}{\sqrt{x}}[\log(x ) + \mathcal{O}(1)].
  \end{equation}
  To obtain the expansion of the t matrix at $T=0$ at large $\omega$
  we use $G(x \to 0)$ in Eq.~\ref{resultT}. Recalling that $T^* = 4
  \pi h^2$, and writing $\log x = \log (\omega x) - \log \omega$ and $y = \omega x$, Eq.~\ref{resultT}  becomes
  \begin{eqnarray}
  t=\frac{1}{2}+ \frac{\sqrt{T^* / \omega}}{(2 \pi)^{3/2}} {\rm{Im}} \int_{-\infty}^\infty dy ~e^{2 i y} \frac{\log y - \log \omega}{\sqrt{y}}.
  \end{eqnarray}
  Note that the $y$ integral should be made symmetrically around the
  branch cut, as described in Sec.\ref{friedel}. One then obtains Eq.\ref{asymptoticT} with
  \begin{eqnarray}
  \beta' &=& -\frac{1+\pi}{\sqrt{2} \pi^{3/2}} \int_0^\infty dy ~\frac{\sin(2y) \log y}{\sqrt{y}}, \nonumber \\
  \delta &=& -\frac{1}{\sqrt{2} \pi^{3/2}} \int_0^\infty dy ~\frac{\sin(2y) }{\sqrt{y}},
  \end{eqnarray}
  with the numerical values of these integrals announced below
  Eq.~\ref{asymptoticT}. At finite temperature and $\omega=0$ we use
  Eq.~\ref{pt} in Eq.~\ref{factorGising}. Taking the limit $\omega \to
  0$ of Eq.~\ref{resultT} and defining $y=x/\beta$, we obtain
  Eq.\ref{asymptoticw} with 
  \begin{equation}
  \beta'' = \frac{4 \sqrt{2}}{\pi} \int_{-\infty}^\infty dy  ~\frac{y K\left(\frac{1-\coth(2 \pi y)}{2} \right)}{\sinh (2 \pi y)} .
  \end{equation}


\section{Linear relations between quadratic forms for original fermions and Majorana fermions}
\label{quadratic}

There are $8 \times 7 = 28$ independent quadratic forms involving
$\psi^\dagger_{\sigma\alpha} $ and $\psi_{\sigma'\alpha'}$, which
together comprise the generators of the $SO(8)$ symmetry group.
These generators are linearly related to the $28$ quadratic forms of
the MFs $\chi_j^A$. In this appendix we gather and re-derive some of these
relations, which can also be found in e.g. Ref.~\onlinecite{maldacena}.

First we define a convention relating the Klein factors $F_A$ for the new
fermions $\psi_A$, to the Klein factors $F_{\sigma\alpha}$ for the
original fermions $\psi_{\sigma\alpha}$. The relations are fully
determined by\cite{ZarandvonDelft}
\begin{equation}
\label{FF}
F_X^\dagger F_s^\dagger = F^\dagger_{\uparrow L} F_{\downarrow
  L},~~~F_X F_s^\dagger = F^\dagger_{\uparrow R} F_{\downarrow R},~~~
F_X^\dagger F_f^\dagger=F^\dagger_{\uparrow L} F_{\uparrow R},
\end{equation}
and by the anticommutation relations~\cite{ZarandvonDelft} $\{ F_A,
F_B \} =2 \delta_{AB}$, $F_A F_A^\dagger =F_A^\dagger F_A= 1$.

We now consider the instructive example of the operator $i \chi_1^f
\chi_1^X$, and use Eqs.~\ref{refermionization} and \ref{MFdecomp} to
relate it to a quadratic term involving the original fermions:
\begin{equation}
\begin{split}
i \chi_1^f \chi_1^X &= \frac{i}{2}  (\psi^\dagger_f+ \psi_f)(\psi^\dagger_X+ \psi_X) = \frac{i}{2}  (\psi^\dagger_f+ \psi_f)\psi^\dagger_X+ \text{H.c.} \\
&=\frac{i}{2} F_f^\dagger F_X^\dagger e^{i \phi_f + i \phi_X} +\frac{i}{2} F_f F_X^\dagger e^{-i \phi_f + i \phi_X}+\text{H.c.}
\end{split}
\end{equation}
Using Eqs.~\ref{FF} and \ref{linear} we obtain,
\begin{equation*}
i \chi_1^f \chi_1^X
=-\frac{i}{2} F^\dagger_{\uparrow L} F_{\uparrow R}   e^{i
  \phi_{\uparrow L} - i \phi_{\uparrow R}} -\frac{i}{2}
F^\dagger_{\downarrow R} F_{\downarrow L} e^{i \phi_{\downarrow R} -i
  \phi_{\downarrow L} }+\text{H.c.}
\end{equation*}
Finally using the bosonization formula Eq.~\ref{bosonization} we have
\begin{equation*}
i \chi_1^f \chi_1^X
=-\frac{i}{2} \psi^\dagger_{\uparrow L} \psi_{\uparrow R}-\frac{i}{2}
\psi^\dagger_{\downarrow R} \psi_{\downarrow L} + \text{H.c.}= \psi^\dagger \frac{\tau^2 \sigma^3}{2}  \psi.
\end{equation*}
In a similar fashion, all of the relations between quadratic forms can
be determined. Conserved currents in the 2CK and 2IK models can be
expressed in terms of the original fermions or the MFs, and the
relations between them are needed for our generalization of the
crossover t matrix to arbitrary perturbation, as considered in
Sec.~\ref{gengf}.  The conserved currents of the 2CK model are,
\begin{equation*}
\label{Jchi1}
\begin{split}
{\rm{charge}}:~~~J
&=\frac{1}{2} \psi^\dagger \psi = i \chi_2^c \chi_1^c  \\
{\rm{spin}}:~~~\vec{J}_s&=\frac{1}{2} \psi^\dagger \vec{\sigma} \psi = -i( \chi_2^s \chi_1^X,\chi_1^X \chi_1^s, \chi_1^s \chi_2^s )  \\
{\rm{flavor}}:~~~\vec{J}_f&=\frac{1}{2} \psi^\dagger \vec{\tau} \psi =
(-i \chi_1^f \chi_2^X, i\chi_2^X \chi_2^f,i \chi_2^f \chi_1^f ).
\end{split}
\end{equation*}
Equivalently, one can define a 3-component spin vector $\vec{\chi}_s =
(\chi_1^s,\chi_2^s,\chi_1^X)$ and flavor vector $\vec{\chi}_f =
(\chi_2^f,-\chi_1^f,-\chi_2^X)$ such that,
\begin{equation}
\label{Jchi2}
\vec{J}_s=\frac{-i}{2}  \vec{\chi}_s \times \vec{\chi}_s, ~~~
\vec{J}_f=\frac{-i}{2}  \vec{\chi}_f \times \vec{\chi}_f.
\end{equation}

Furthermore, the 9 spin-flavor current components can be expressed as,
\begin{equation}
\frac{1}{2} \psi^\dagger \sigma^a \tau^b \psi = i (\vec{\chi}_s)^a (\vec{\chi}_f)^b, ~~~(a,b=x,y,z).
\end{equation}

Thus, the decomposition of the 2CK model into $U(1) \times SU(2)_2
\times SU(2)_2$ charge, spin and flavor sectors can be understood
also in terms of MFs.

In the 2IK model, there is no flavor $SU(2)$ symmetry since each
channel couples to a different impurity.  However, one can make use of
the $SU(2)$ total spin current $\vec{J}_s$, as well as $SU(2)$ isospin
currents for each channel $\vec{I}_L$ and $\vec{I}_R$, where
\begin{equation}
I_\alpha^z = \frac{1}{2} \sum_{\sigma}\psi^\dagger_{\sigma\alpha}  \psi_{\sigma\alpha},
~~~~I_{\alpha}^- = \psi_{\uparrow \alpha} \psi_{\downarrow \alpha}.
\end{equation}
In terms of MFs, we have
\begin{equation}
\begin{split}
I_L^z+I_R^z & =J=i \chi_2^c \chi_1^c,  \\
I_L^z-I_R^z & =J_f^z=i \chi_2^f \chi_1^f,
\end{split}
\end{equation}
and
\begin{equation}
\begin{split}
I_L^x+I_R^x & =i \chi_1^f \chi_2^c,\\
I_L^x-I_R^x & =i \chi_2^f \chi_1^c.
\end{split}
\end{equation}
Hence one can understand the conformal embedding of the 2IK model as a
$SU(2)_2 \times SU(2)_1 \times SU(2)_1 \times \rm{Ising}$
decomposition into total spin, left/right channel isospin and Ising
sectors. 3 of the 8 MFs represent the total spin sector; 4 represent
the isospin symmetry sectors (the charge and flavor MFs); and the
remaining MF, $\chi_2^X$, is associated with the Ising model (and
restores the total central charge $c=4$).




\begin{thebibliography}{78}
\bibitem[{not()}]{note:convention}
\bibinfo{note}{A different convention is employed in
  Ref.~\onlinecite{SelaMitchellFritz}, which corresponds to a trivial
  relabelling of the undetermined constants $c_1\rightarrow \sqrt{c_T}$,
  $c_V\rightarrow \sqrt{c_T}c_V$ and $c_B\rightarrow -\sqrt{c_T}c_B$.}
\end{thebibliography}
\end{document}